\documentclass[preprintnumbers,amsmath,amssymb,floatfix,10pt,prd,twocolumn,superscriptaddress,nofootinbib]{revtex4-2}
\usepackage{latexsym}
\usepackage{physics}
\usepackage{epsfig}
\usepackage{epstopdf}
\usepackage{graphicx}
\usepackage{amssymb}
\usepackage{amsmath}
\usepackage{amsfonts}
\usepackage{subfigure}
\usepackage{dcolumn}
\usepackage{bm}
\usepackage{color}
\usepackage{comment}
\usepackage[utf8]{inputenc}
\usepackage[dvipsnames]{xcolor}
\usepackage{hyperref}
\hypersetup{colorlinks=true,linkcolor=blue,citecolor=red,urlcolor=magenta}
\usepackage{orcidlink}
\usepackage{lipsum}

\begin{document}
\title{\bf Rotating Einstein-Maxwell-Dilaton Black Hole as a Particle Accelerator}

\author{Muhammad Ali Raza \orcidlink{0009-0001-1281-8702}}
\email{maliraza01234@gmail.com}
\affiliation{Department of Mathematics, COMSATS University Islamabad, Lahore Campus, Lahore, Pakistan}

\author{Sehar}
\email{seharimran958@gmail.com}
\affiliation{Department of Mathematics, Division of Science and Technology, University of Education, Lahore, Pakistan.}

\author{M. Azam}
\email{azam.math@ue.edu.pk}
\affiliation{Department of Mathematics, Division of Science and Technology, University of Education, Lahore, Pakistan.}

\author{M. Zubair \orcidlink{0000-0003-2227-788X}}
\email{mzubairkk@gmail.com;drmzubair@cuilahore.edu.pk}
\affiliation{Department of Mathematics, COMSATS University Islamabad, Lahore Campus, Lahore, Pakistan}
\affiliation{National Astronomical Observatories, Chinese Academy of Sciences,
Beijing 100101, China}

\author{Francisco Tello-Ortiz$^{*}$ \orcidlink{0000-0002-7104-5746}}
\email{francisco.tello@ufrontera.cl}
\affiliation{Departamento de Ciencias Físicas, Universidad de La Frontera, Casilla 54-D, 4811186 Temuco, Chile.}

\begin{abstract}
Similar to particle accelerators, black holes also have the ability to accelerate particles, generating significant amounts of energy through particle collisions. In this study, we examine the horizon and spacetime structures of a rotating black hole within the framework of Einstein-Maxwell-Dilaton gravity. Additionally, we extend the analysis to explore particle collisions and energy extraction near this black hole using the Ba\~{n}ados-Silk-West mechanism. Our findings reveal that the mass and angular momentum of the colliding particles significantly influence the center of mass energy, more so than the parameters of the black hole itself. Furthermore, we apply the  Ba\~{n}ados-Silk-West mechanism to massless particles, particularly photons, while disregarding their intrinsic spin in plasma; an aspect that has not been previously explored. The  Ba\~{n}ados-Silk-West mechanism cannot be directly applied, as the refractive index condition only permits photon propagation, meaning that massive particles in vacuum cannot be included in this study. We derive the propagation conditions for photons and analyze photon collisions by treating them as massive particles in a dispersive medium. The impact of the plasma parameter on the extracted center of mass energy is also examined. Our results show that the plasma parameter has a relatively weak and unchanged effect on the center of mass energy across all cases, indicating that energy losses due to friction within the medium are a contributing factor.

\end{abstract}
\maketitle
\date{\today}

\section{Introduction}\label{S1}
A black hole can be conceptualized as a particle accelerator due to the immense gravitational forces near its event horizon. A significant amount of energy can be extracted from particle collisions occurring in the strong gravitational field around black holes. One might question the existence of particles within the black hole's field; however, due to the influence of gravity and accretion disks, matter likely exists in the vicinity of black holes, where numerous particle collisions can occur. The energy extracted from these particle collisions is typically measured in terms of the center of mass energy (CME) for the system, rather than the energies of the individual constituents. This is a purely relativistic phenomenon and plays a crucial role in high-energy astrophysics.
\textcolor{blue}{When a black hole has rotation, the spacetime structure notably changes from the static case, creating a region known as the ergosphere, within which the phenomenon known as frame dragging becomes significant. This effect inevitably drags any particle or radiation within this region in the direction of the black hole’s rotation, introducing important additional dynamics for the analysis of particle collisions near these compact objects \cite{Bardeen:1972fi}.
Moreover, considering modified gravity theories such as Einstein-Maxwell-Dilaton (EMD) gravity is particularly relevant since these theories naturally emerge from fundamental contexts like string theory or supergravity models. Specifically, the additional scalar field known as the dilaton not only modifies the structure of the black hole itself but also significantly influences how matter and radiation interact in its vicinity. This influence is critical because coupled scalar fields can substantially alter horizon formation, spacetime geometry near the black hole, and extreme energetic processes such as high-energy particle collisions \cite{Garfinkle:1990qj,Gibbons:1987ps,Sen:1992ua}.
Additionally, explicitly considering a dispersive medium, such as plasma, becomes essential since plasma is commonly found around real black holes. Thus, studying photon propagation and collisions in these dispersive media—effectively treated as massive particles due to plasma dispersive properties—allows modeling realistic conditions under which emitted radiation and observable processes in extreme astrophysical environments could be notably influenced.
By investigating these phenomena through some mechanism (see below), this study deepens the theoretical understanding of particle dynamics in intense gravitational fields and offers quantitative predictions potentially testable by future high-precision astronomical observations, thereby strengthening the connection between theory and observation in relativistic astrophysics.}

Baushev \cite{2009IJMPD..18.1195B} was the first to propose that a black hole could act as a particle accelerator. He examined the Schwarzschild black hole and discovered that particles near the black hole could collide, resulting in the production of a substantial amount of energy. However, his study did not provide a formal or detailed analysis of particle collisions in the vicinity of a black hole. Later on, Ba\~{n}ados et al. \cite{PhysRevLett.103.111102} developed a formal procedure to calculate the energy generated by particle collisions near a black hole. They considered two uncharged particles of equal mass colliding near a Kerr black hole and found a significant amount of CME. They also extended the analysis to the case of a Schwarzschild black hole. This method later became known as the Ba\~{n}ados-Silk-West (BSW) mechanism. However, Berti et al. \cite{PhysRevLett.103.239001} proposed that such collisions are less likely to occur in nature, by discussing the BSW mechanism. They found that the upper bound on the spin of the Kerr black hole, along with radiation losses, imposes an upper limit on the CME. Additionally, Jacobson and Sotiriou \cite{PhysRevLett.104.021101} highlighted the physical limitations of such high-energy collisions. In the same year, Lake \cite{PhysRevLett.104.211102,PhysRevLett.104.259903} suggested that classical black holes are internally unstable due to a divergent CME for two neutral, massive colliding particles at the Cauchy horizon of a Kerr black hole. Grib and Pavlov \cite{2010JETPL..92..125G,2011GrCo...17...42G,2011APh....34..581G} investigated particle collisions and observed a high CME resulting from scattering in both non-extremal and extremal Kerr black holes. Harada and Kimura \cite{PhysRevD.83.024002} derived an analytic, general relation to calculate the center-of-mass energy (CME) for the collision of two particles with equal masses near the event horizon of a Kerr black hole. This relation was applied to two distinct collision scenarios: in the first, a particle traveling from the innermost stable circular orbit collided with another particle closer to the event horizon, while in the second, both particles were in the innermost stable circular orbit. They identified the maximum CME limits for extremal Kerr black holes. Based on the BSW mechanism some notable works are also given in the Refs. \cite{PhysRevD.82.083004,PhysRevD.82.103005,2010JHEP...12..066W,PhysRevD.83.084041,PhysRevD.85.024020,PhysRevD.88.027505,PhysRevD.88.124001,2014EPJC...74.2759C,2015EPJC...75...24J} and some of the recent advances in particle collision theory and energy extraction from the center of mass frame by using the BSW mechanism is also given in the Refs. \cite{doi:10.1142/S0218271821501108,2023EPJP..138..846N,2024EPJC...84..203K,doi:10.1142/S0218271824500445}.

Due to the immense gravity near black holes, matter is drawn in and trapped within specific regions around them. Additionally, under the influence of Hawking radiation, black holes emit energy in the form of radiation \cite{hawking1974black,1975CMaPh..43..199H}. Consequently, a significant number of collisions and interactions can occur in the gravitational field of black holes, raising the temperature enough to transform matter into plasma. Recently, the Event Horizon Telescope collaboration captured an image of the supermassive black hole M87* \cite{collaboration2019first}, and a few months later, the same collaboration found evidence of a magnetic field around it, suggesting the presence of a plasma medium surrounding black holes \cite{akiyama2021first}. Black holes surrounded by plasma can also be viewed as particle accelerators, and the CME generated from particle collisions in such environments can be studied. However, the BSW mechanism limits this kind of analysis because it assumes that the colliding particles have a non-zero rest mass. While one might consider massive particles, the condition for their propagation in plasma cannot be determined using the refractive index relation, which is only applicable to photons. To address this issue, one could assume the motion occurs entirely within the plasma, allowing photons to behave as timelike particles with non-zero rest mass. In this way, the propagation conditions for such particles can be determined, and the BSW mechanism can be applied.

Motivated by this, we investigate the particle collision and CME in vacuum and plasma around the rotating black hole in Einstein-Maxwell-Dilaton (EMD) gravity \cite{doi:10.1142/S0218271819500639} whose static counterpart is given in Refs. \cite{PhysRevD.43.3140,1988NuPhB.298..741G}. General Relativity is a well-established theory today, following over a century of progress. However, there are still unanswered questions that it has yet to explain. One such question is the justification for the accelerated expansion of the universe, for which the leading candidate is dark energy. Various modified theories have been proposed to better understand the nature of dark energy. One such theory arises from coupling General Relativity with Maxwell's electrodynamics and a phantom dilatonic field under different conditions. The black hole under consideration is derived by assuming the action outlined in the Refs. \cite{PhysRevD.43.3140,1988NuPhB.298..741G}, in which a scalar (dilaton) field $\phi$ is coupled to the Maxwell's field by a factor $e^{-2\alpha\phi}$, with $\alpha$ being the coupling constant. For $\alpha=1$, the theory is restricted to string theory at low energy. However, it is useful to keep $\alpha$ as a parameter. For $\alpha=0$, the theory reduces to Einstein-Maxwell electrodynamics generating the Reissner-Nordstr\"{o}m black hole solution. Therefore, the EMD gravity and the corresponding black hole solution derived from this theory hold significant importance.

The paper is organized as: In Sec. \ref{S2}, the static black hole metric is presented, following with its rotating counterpart along with horizon structure and ergosphere in Sec. \ref{S3}. Section \ref{S4} focuses on timelike geodesics and particle orbits via effective potential. The Secs. \ref{S5} and \ref{S6} comprise the discussion on CME in non-plasma (vacuum) and plasma media, respectively. Finally, the paper is concluded in Sec. \ref{S7}.

\section{The Static Black Hole}\label{S2}
The static black hole solution in EMD gravity is derived by considering the action that describes the coupling of General Relativity with the dilaton and Maxwell fields, which is mathematically expressed as \cite{PhysRevD.43.3140,1988NuPhB.298..741G}
\begin{eqnarray}
S=\int\sqrt{-g}\left[R-2\left(\nabla\phi\right)^2-e^{-2\alpha\phi}F_{\mu\nu}F^{\mu\nu}\right]d^4x, \label{1}
\end{eqnarray}
where, $g$ is the determinant of the metric tensor $g_{\mu\nu}$. Moreover, $R$ is the scalar curvature, and the coupling between the dilaton field $\phi$ and the Maxwell's tensor $F_{\mu\nu}$ is defined by the parameter $\alpha$. Interestingly, the sign of $\alpha$ can be changed equivalently by changing the sign of $\phi$. Therefore, for not losing the generality, one may consider $\alpha\geq0$. If we consider $\alpha=0$, the action (\ref{1}) is reduced to the well-known Einstein-Maxwell action together with a minimally coupled scalar field. Moreover, when $\alpha=1$, this action is part of string theory at the low-energy limit. The equations of motion determined from (\ref{1}) are as follows:
\begin{eqnarray}
\nabla_\mu\left(e^{-2\alpha\phi}F^{\mu\nu}\right)&=&0, \label{2}\\
2\nabla^2\phi+\alpha e^{-2\alpha\phi}F_{\mu\nu}F^{\mu\nu}&=&0, \label{3}\\
R_{\mu\nu}-\nabla_\mu\phi\nabla_\nu\phi-\frac{4F_{\mu\sigma}F^\sigma_\nu-g_{\mu\nu}F_{\sigma\beta}F^{\sigma\beta}}{2e^{2\alpha\phi}}&=&0. \label{4}
\end{eqnarray}
The general static spherically symmetric line element
\begin{eqnarray}
ds^2=-f(r)dt^2+\frac{dr^2}{f(r)}+h(r)\left(d\theta^2+\sin^2\theta d\phi^2\right) \label{5}
\end{eqnarray}
admits the solution for the metric functions $f(r)$ and $h(r)$ \cite{PhysRevD.43.3140,1988NuPhB.298..741G}	given as
\begin{eqnarray}
f(r)&=&\left(1-\frac{r_1}{r}\right)\left(1-\frac{r_2}{r}\right)^\frac{1-\alpha^2}{1+\alpha^2}, \label{6}\\
h(r)&=&r^2\left(1-\frac{r_2}{r}\right)^\frac{2\alpha^2}{1+\alpha^2}, \label{7}
\end{eqnarray}
where, the black hole mass $M$ and the charge $Q$ are related to the parameters $r_1$ and $r_2$ by
\begin{eqnarray}
M&=&\frac{r_1}{2}+\left(\frac{1-\alpha^2}{1+\alpha^2}\right)\frac{r_2}{2}, \label{8}\\
Q^2&=&\frac{r_1r_2}{1+\alpha^2}. \label{9}
\end{eqnarray}
The Maxwell and dilaton fields are defined by
\begin{eqnarray}
F_{tr}&=&\frac{Q}{r^2}, \label{10}\\
e^{2\phi}&=&\left(1-\frac{r_2}{r}\right)^\frac{2\alpha}{1+\alpha^2}, \label{11}
\end{eqnarray}
respectively. The radii $r_1$ and $r_2$, in terms of the mass and charge, may be obtained by inverting Eqs. (\ref{8}) and (\ref{9}) as
\begin{eqnarray}
r_1&=&M+\sqrt{M^2-\left(1-\alpha^2\right)Q^2}, \label{12}\\
r_2&=&\frac{1+\alpha^2}{1-\alpha^2}\left(M-\sqrt{M^2-\left(1-\alpha^2\right)Q^2}\right). \label{13}
\end{eqnarray}
The Eqs. (\ref{12}) and (\ref{13}) are quadratic in nature, therefore, the signs corresponding to positive solutions have been chosen. The metric (\ref{5}) may only be considered to give physical solution if $\left(1-\alpha^2\right)Q^2\leq M^2$ in Eqs. (\ref{12}) and (\ref{13}) because the radius is always a real number. By considering $\alpha\geq1$, this condition holds automatically. However, if $\alpha<1$, then the charge is bounded above as
\begin{equation}
Q^2\leq\frac{M^2}{1-\alpha^2}. \label{14}
\end{equation}
There may exist a naked singularity in the spacetime if the condition (\ref{14}) is satisfied. The solution simplifies to the Reissner-Nordstr\"{o}m metric for $\alpha=0$ with horizons $r_1$ and $r_2$, whereas, a point singularity is defined at $r=0$. In order to avoid a naked singularity, we require $r_1>r_2$. This is because, for every $\alpha>0$, the geometry at $r_2$ is singular, while the horizons remain at $r_1$ and $r_2$. In terms of the mass and charge, this corresponds to the condition
\begin{equation}
Q^2\leq\left(1+\alpha^2\right)M^2 \label{15}
\end{equation}
for the existence of an event horizon.

\section{The Rotating Black Hole}\label{S3}
We know that the rotating black hole metrics in EMD gravity only correspond to closed form for $\alpha=0$ and $\alpha=\sqrt{3}$ \cite{PhysRevD.46.1340}. The case when $\alpha=0$, one gets the Kerr-Newman black hole metric, and $\alpha=\sqrt{3}$ generates the rotating black hole metric in Kaluza-Klein theory. For $\alpha=1$, the application of the Newman-Janis algorithm to the static metric (\ref{5}) generates a pre-existing Kerr-Sen solution \cite{Yazadjiev:1999ce,PhysRevLett.69.1006}. Surprisingly, this Kerr-Sen solution does not satisfy the field equations (\ref{2})-(\ref{4}) unless an axion field is included in the action (\ref{1}). Azreg-A\"{i}nou modified the Newman-Janis algorithm \cite{Azreg-Ainou:2014aqa,PhysRevD.90.064041} to deal with such issues. In particular, this algorithm is considered to obtain the rotating black hole metric in EMD gravity in Ref. \cite{doi:10.1142/S0218271819500639} for arbitrary values of $\alpha$ which is given as
\begin{eqnarray}
ds^2&=&-\frac{H\Delta(r)}{\Sigma}dt^2+\frac{\Sigma\sin^2\theta}{H}\left(d\phi-\frac{a\sigma}{\Sigma}dt\right)^2+\frac{H}{\Delta(r)}dr^2\nonumber\\&&+Hd\theta^2, \label{16}
\end{eqnarray}
where,
\begin{align}
H&=h(r)+a^2\cos^2\theta, \label{17}\\
\Delta(r)&=f(r)h(r)+a^2=r^2-\left(r_1+r_2\right)r+r_1r_2+a^2, \label{18}\\
\sigma&=h(r)(1-f(r)), \label{19}\\
\Sigma&=(h(r)+a^2)^2-a^2\Delta(r)\sin^2\theta, \label{20}
\end{align}
such that $a$ is the black hole spin. For the above rotating metric, the cross term $G_{r\theta}$ of Einstein tensor vanishes. Therefore, one may regard the rotating metric (\ref{16}) as a solution of the initial field equations. It is because of the fact that the energy-momentum tensor can be written as \cite{Azreg-Ainou:2014aqa,PhysRevD.90.064041}
\begin{equation}
T^{\mu\nu}=\epsilon e^\mu_te^\nu_t+p_re^\mu_re^\nu_r+p_\theta e^\mu_\theta e^\nu_\theta+p_\phi e^\mu_\phi e^\nu_\phi, \label{21}
\end{equation}
where, the vectors $e_r$ and $e_\theta$ in the orthonormal tetrad ($e_t,e_r,e_\theta,e_\phi$) are related to the basis vectors $\partial r$ and $\partial\theta$. Therefore, the gravitational source $T_{\mu\nu}$ described in Eq. (\ref{21}) can be regarded as an imperfect fluid that rotates about $z$-axis \cite{Azreg-Ainou:2014aqa,PhysRevD.90.064041}. The roots of equation $\Delta(r)=0$ define two horizons given by
\begin{equation}
r_{\pm}=\frac{r_1+r_2\pm\sqrt{\left(r_1-r_2\right)^2-4a^2}}{2}. \label{22}
\end{equation}
The rotating black hole in EMD gravity reduces to Kerr-Newman black hole for $\alpha=0$ with a ring singularity at origin and $\theta=\frac{\pi}{2}$, and two horizons at
\begin{equation}
r_{\pm}=M\pm\sqrt{M^2-a^2-Q^2}. \label{23}
\end{equation}
However, for $\alpha>0$, the singularity becomes more complicated as it depends upon $a$ and $\theta$ as well besides $\alpha$, $M$ and $Q$ \cite{doi:10.1142/S0218271819500639}. For a well-defined black hole solution we must have the condition
\begin{equation}
r_1-r_2\geq2|a| \label{24}
\end{equation}
in order to avoid the naked singularity, contrary to $r_1-r_2>0$ as for the non-rotating case. In terms of mass and charge, the condition (\ref{24}) becomes
\begin{figure*}
\centering
\subfigure{\includegraphics[width=0.35\textwidth]{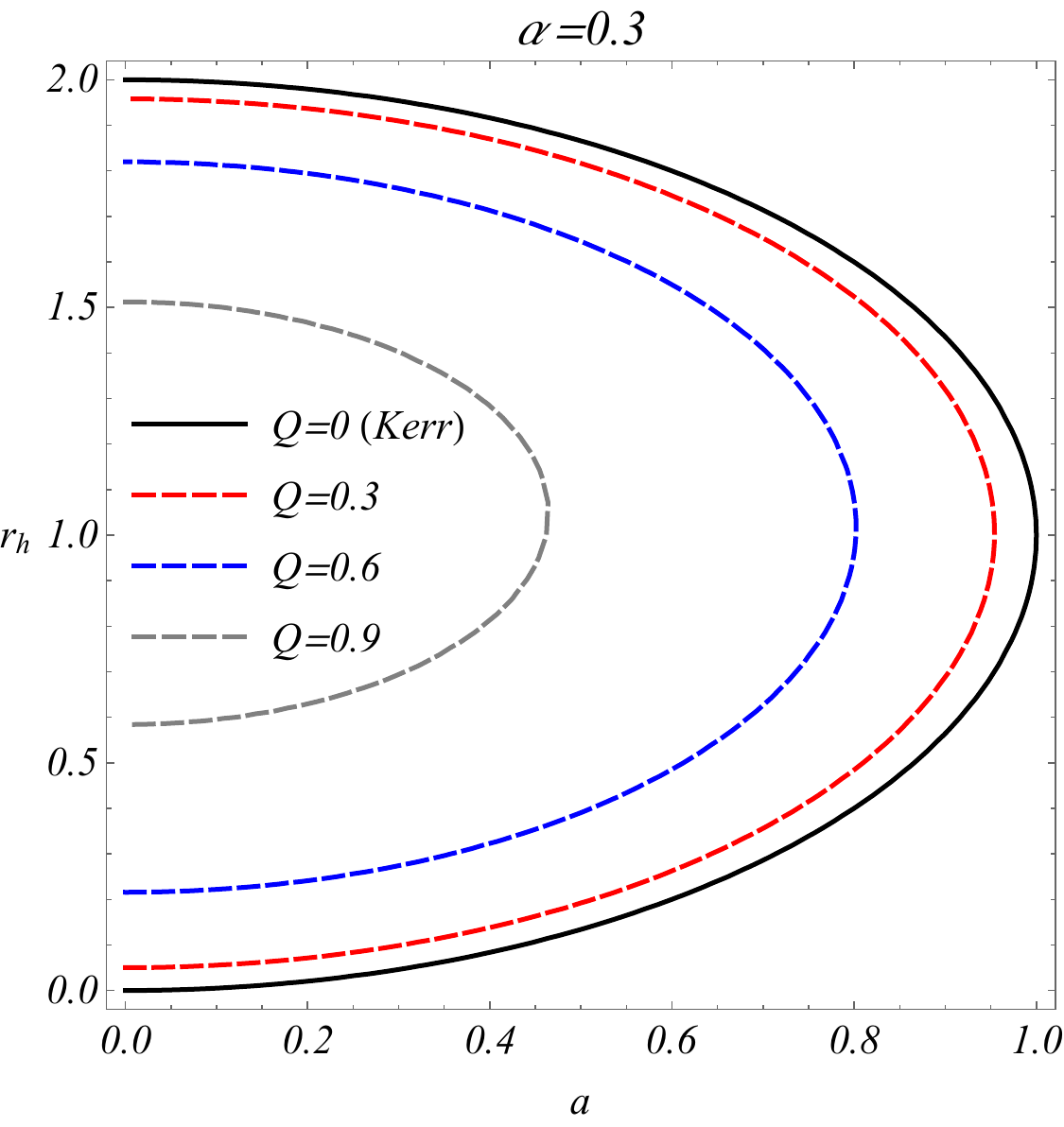}}~~~~~~
\subfigure{\includegraphics[width=0.35\textwidth]{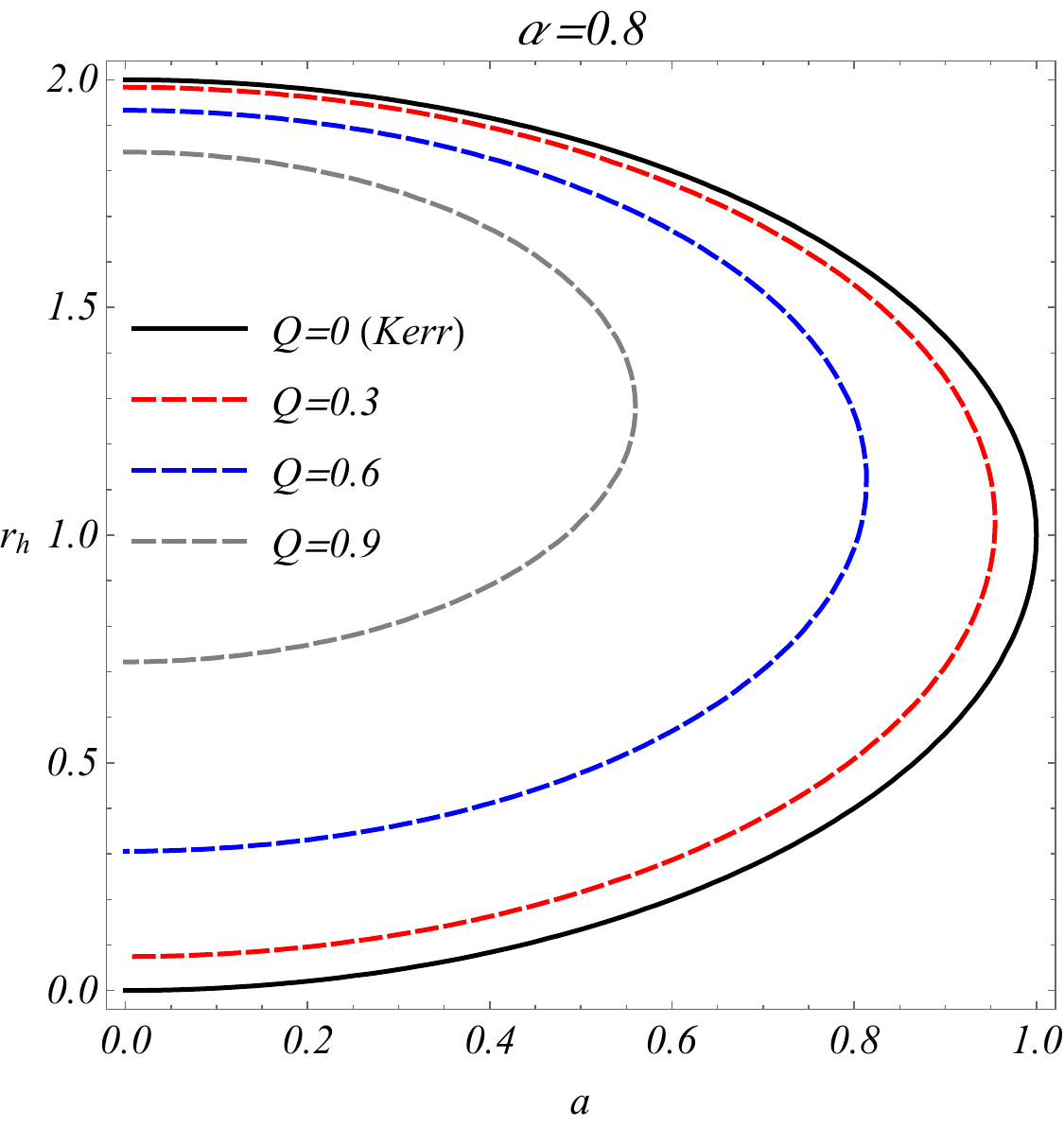}}
\subfigure{\includegraphics[width=0.35\textwidth]{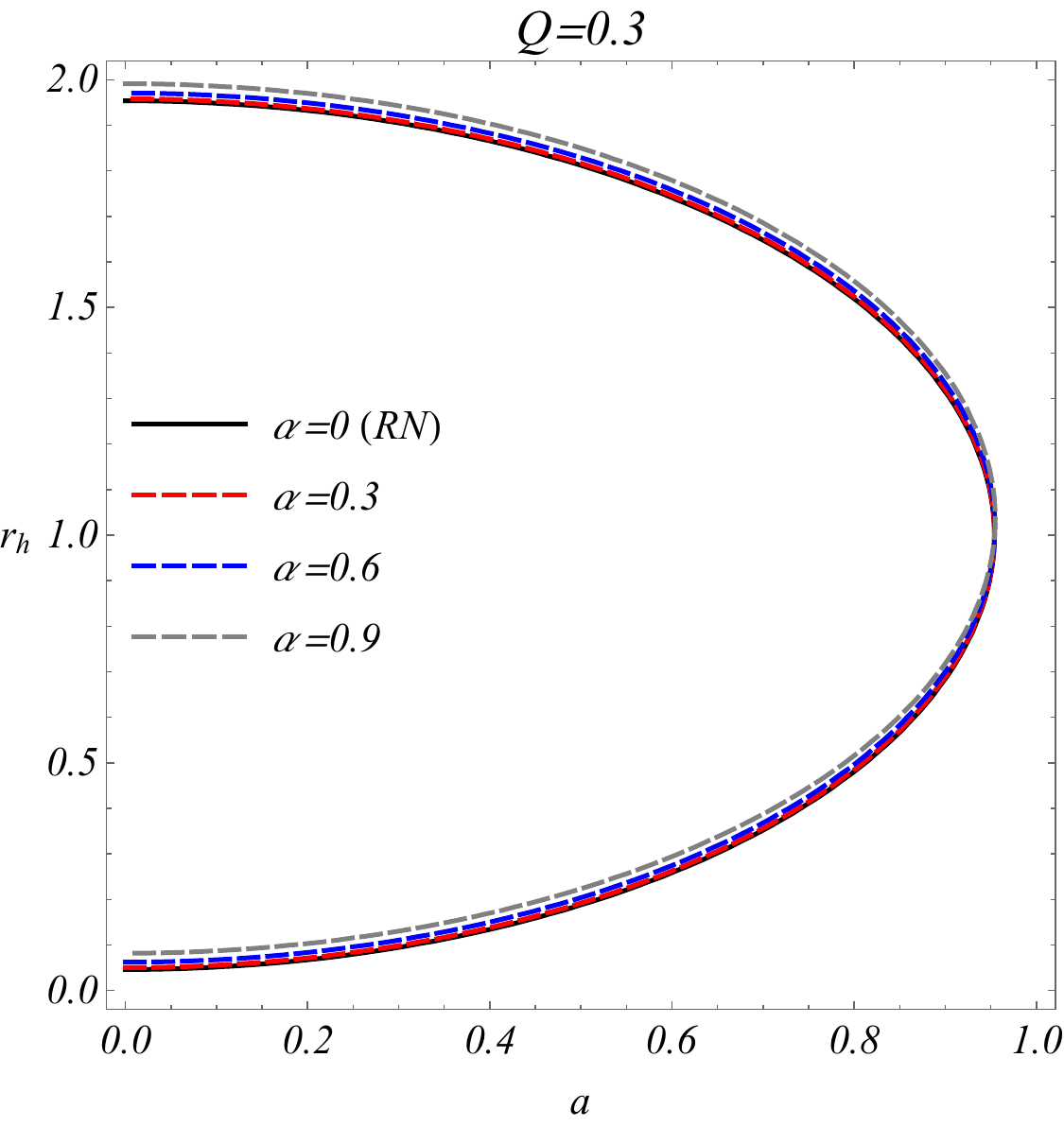}}~~~~~~
\subfigure{\includegraphics[width=0.35\textwidth]{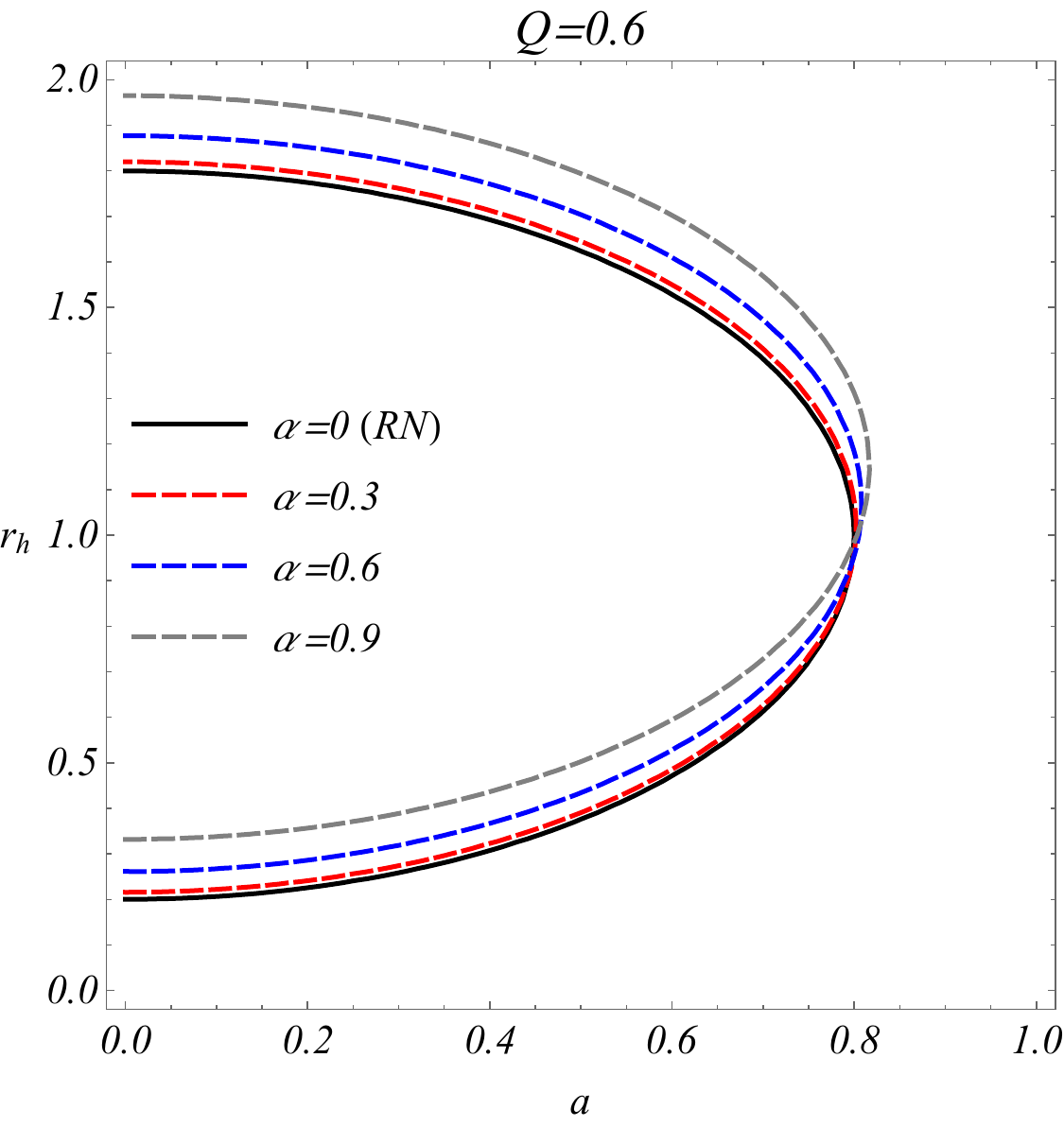}}
\caption{Plots illustrating the horizon structure with respect to $a$ for various values of $\alpha$ and $Q$ with $M=1$. \label{f1}}
\end{figure*}
\begin{equation}
Q^2\leq\left(|a|-M\right)\left[|a|\left(\alpha^2-1\right)-M\left(\alpha^2+1\right)\right]. \label{25}
\end{equation}
If $|a|\leq M$, the condition (\ref{25}) holds for $\alpha\leq1$. Whereas, $\alpha>1$ is not a physically viable interval as it generates a branch of solutions with $|a|\left(\alpha^2-1\right)\geq\left(\alpha^2+1\right)M$.

The event and Cauchy horizons for the spinning black hole (\ref{16}) in EMD gravity are given by the Eq. (\ref{22}). We have examined the horizon structure with respect to spin $a$ for different values of the parameters $\alpha$ and $Q$ keeping the mass $M$ fixed. The behavior of horizon radii are plotted with respect to spin parameter $a$ for various values of $\alpha$ and $Q$ in Fig. \ref{f1}. The upper panel comprise the curves corresponding to $Q$, whereas, the curves in the lower panel correspond to $\alpha$. For all plots, it is observed that the event horizon decreases and the Cauchy horizon increases with respect increase in $a$ up to the extremal value of $a$. In the upper panel, the event horizon decreases with increase in $Q$. However, as $\alpha$ increases, the decrement in event horizon reduces with increase in $Q$. The outermost curves correspond to the case of Kerr black hole with the largest event horizon and extremal spin values as the effect of $\alpha$ vanishes with $Q=0$. A converse behavior is observed in the lower panel, that is, with increase in $\alpha$, the event horizon increases. This rate further increases with increase in $Q$. This show a converse dependence of $\alpha$ and $Q$ on each other. The curves corresponding to $\alpha=0$ show the deviation of the results for Reissner-Nordstr\'{o}m black hole from the rotating black hole in EMD gravity.

The rotating black holes are bounded by the event horizon from inside and a static limit surface as its outer bound. This enclosed region is known as ergosphere, in which, if a particle or an object enters, will move with the spin of the black hole. There is another static limit surface inside the Cauchy horizon. These static limit surfaces are the real and positive roots of the equation $g_{tt}=0$, denoted as $r^+_{sls}$ and $r^-_{sls}$. Since, the Hawking radiation can be studied in this region, the ergoregion is crucial in astrophysics for potential observational studies. The following equation defines the static limit surfaces:
\begin{equation}
g_{tt}=H^2\Delta(r)-\sigma^2a^2\sin^2\theta=0. \label{26}
\end{equation}
To plot the behavior of the static limit surfaces and to illustrate the structure of ergosphere, we consider spheroidal coordinates with transformation equations given as
\begin{figure*}
\centering
\subfigure{\includegraphics[width=0.37\textwidth]{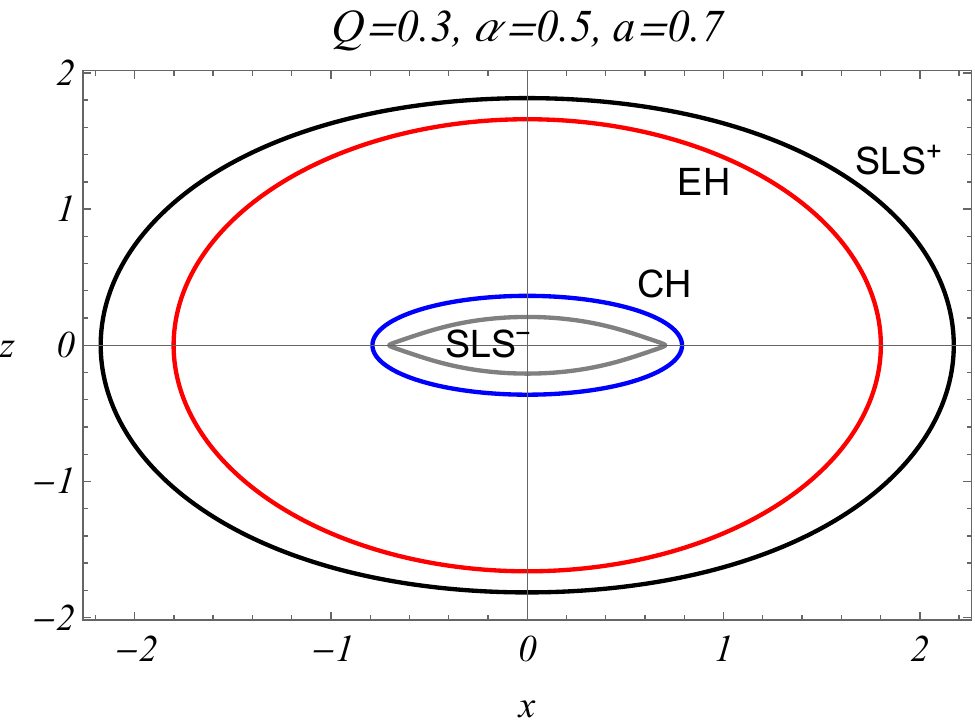}}~~~
\subfigure{\includegraphics[width=0.37\textwidth]{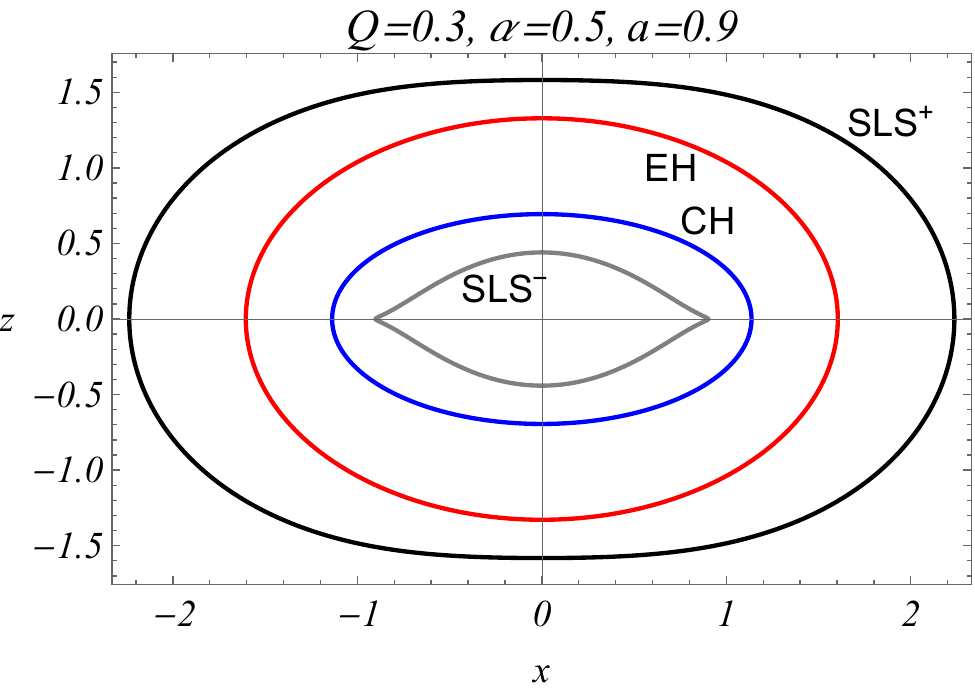}}
\subfigure{\includegraphics[width=0.37\textwidth]{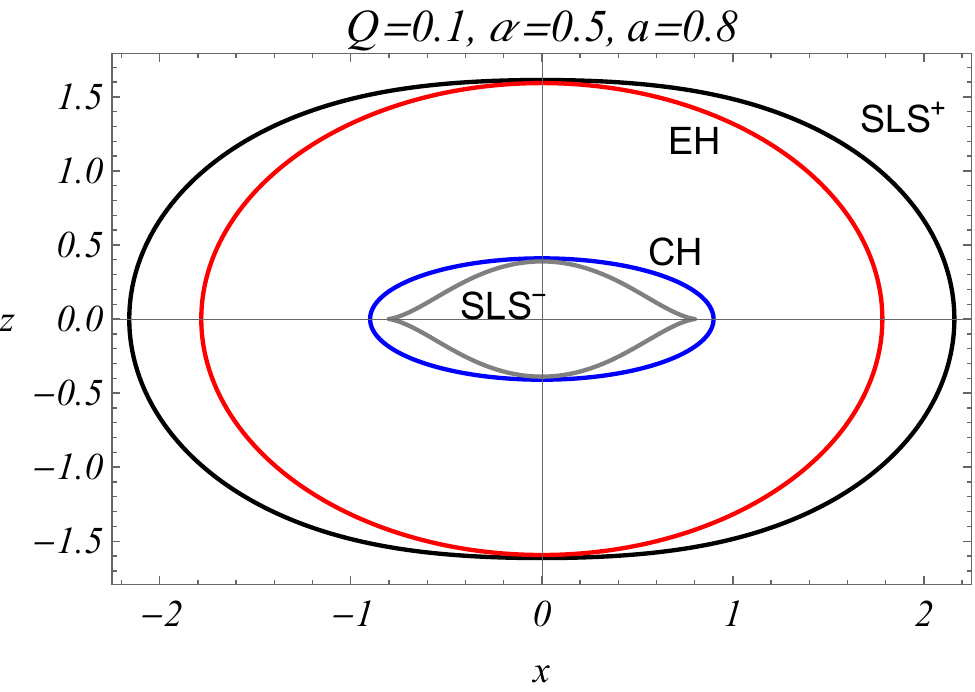}}~~~
\subfigure{\includegraphics[width=0.37\textwidth]{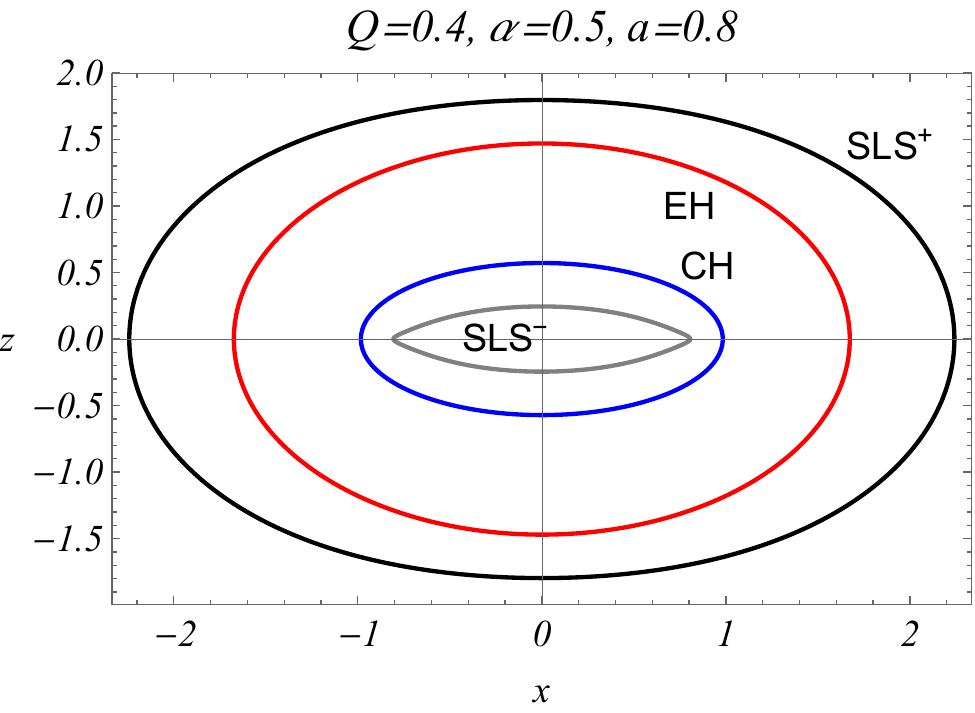}}
\subfigure{\includegraphics[width=0.37\textwidth]{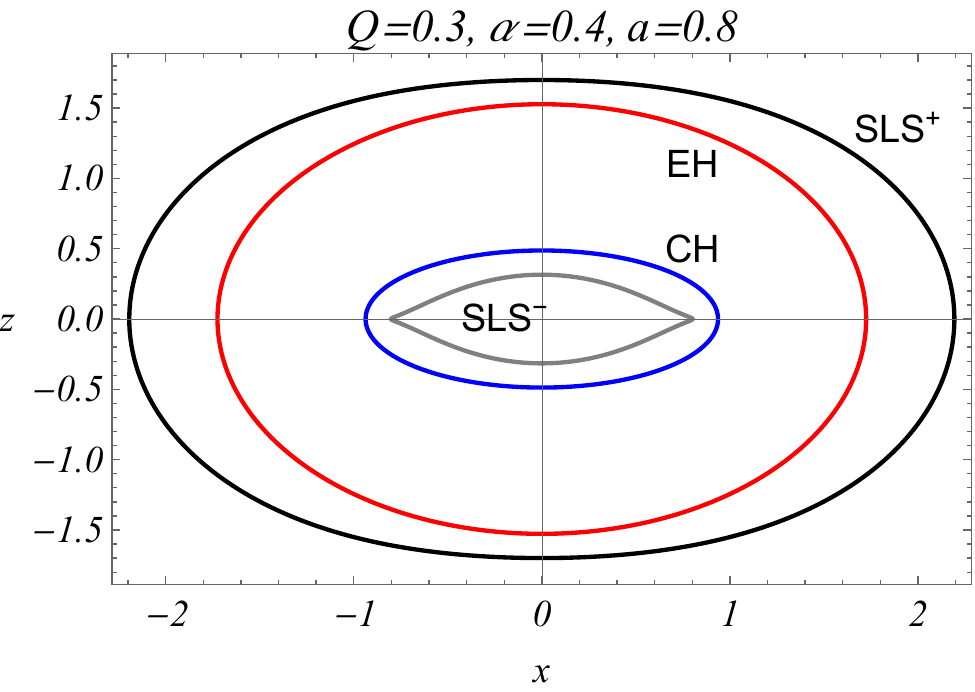}}~~~
\subfigure{\includegraphics[width=0.37\textwidth]{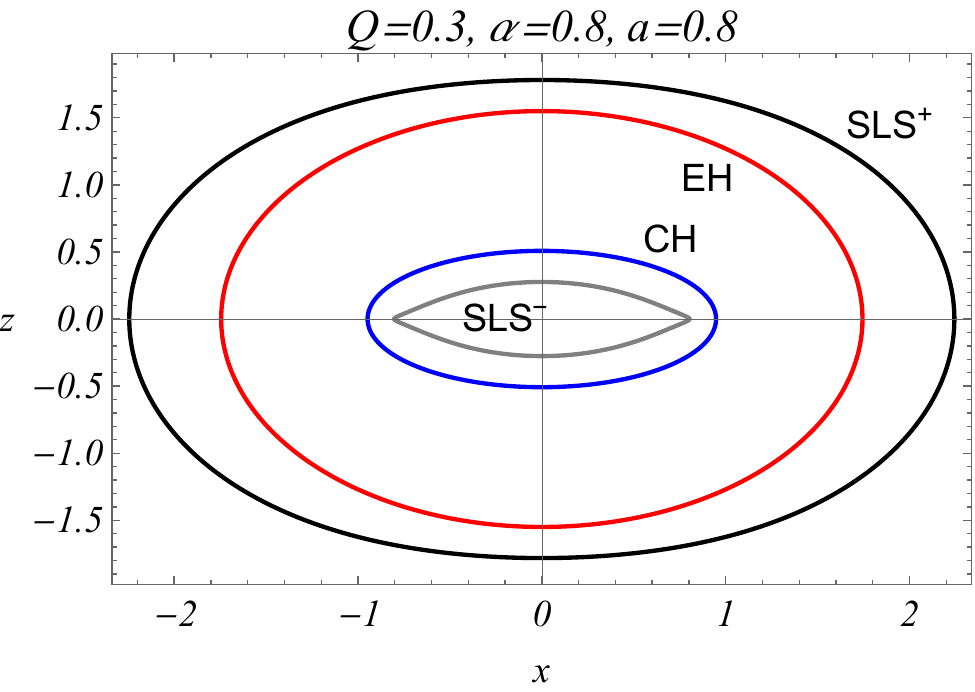}}
\caption{Plots showing the structure for static limit surfaces and horizons for various values of $a$, $\alpha$ and $Q$ with $M=1$. \label{f2}}
\end{figure*}
\begin{eqnarray}
x&=&\sqrt{r^2+a^2}\cos\phi\sin\theta, \label{27}\\ 
y&=&\sqrt{r^2+a^2}\sin\phi\sin\theta, \label{28}\\
z&=&r\cos\theta. \label{29}
\end{eqnarray}
The spacetime structure of the rotating black hole in EMD gravity comprising the static limit surfaces and the horizons in spheroidal coordinate system is shown in Fig. \ref{f2} for various values of $a$, $\alpha$ and $Q$ with $M=1$. The spacetime diagrams are plotted in $x$-$z$ plane with $y$-axis pointing vertically outwards. In the top panel, $Q$ and $\alpha$ has been kept fixed and $a$ is varied. It clearly shows that the size of ergosphere increases with increase in spin. Whereas, for fixed $a$ and $\alpha$ in the middle panel, the ergosphere increases with increase in $Q$. In the lower panel, there is no significant variation in the ergosphere with respect to increase in $\alpha$ for the given values of $a$ and $Q$. It is worth noticing that there also exists an ergoregion at the poles as well which suggests the precession of the black hole.

\section{Particle Orbits and Effective Potential}\label{S4}
\begin{figure*}
\centering
\subfigure{\includegraphics[width=0.4\textwidth]{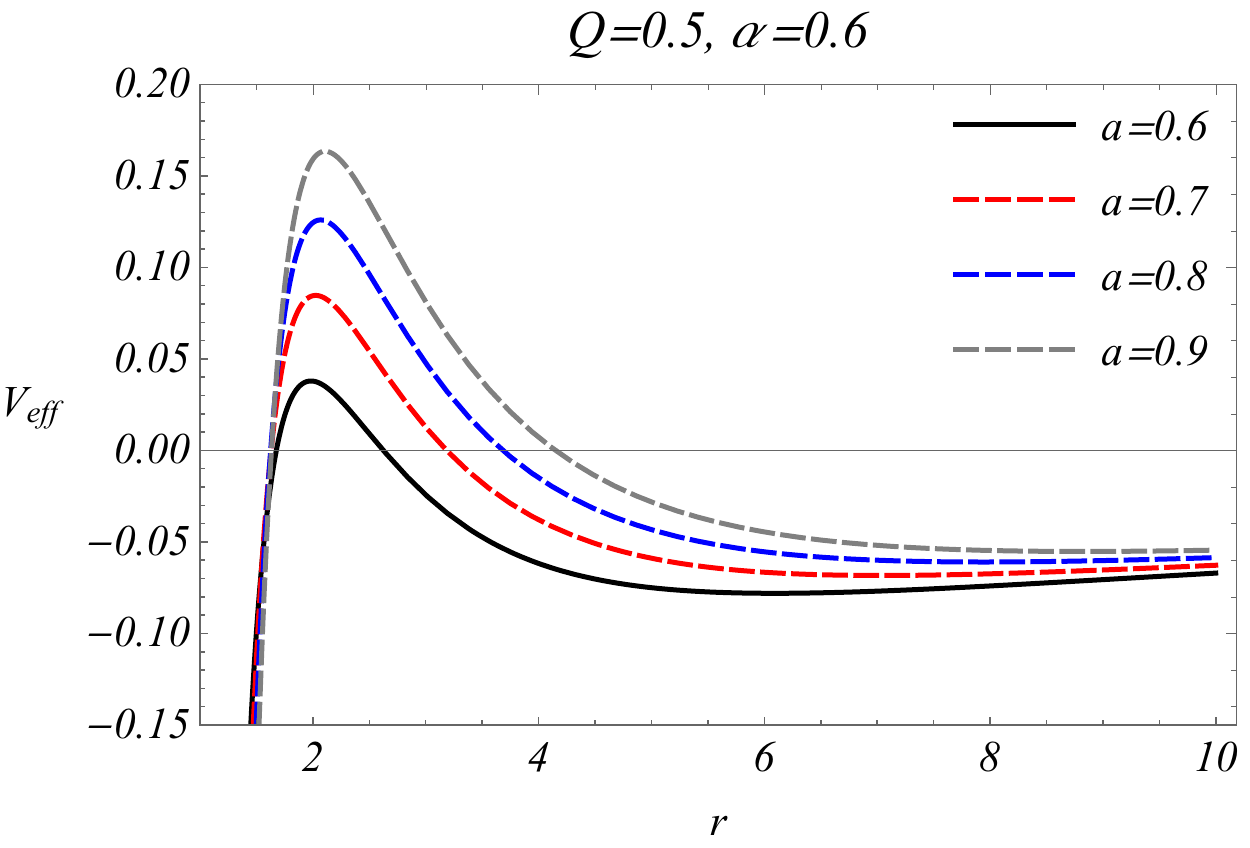}}~~~
\subfigure{\includegraphics[width=0.4\textwidth]{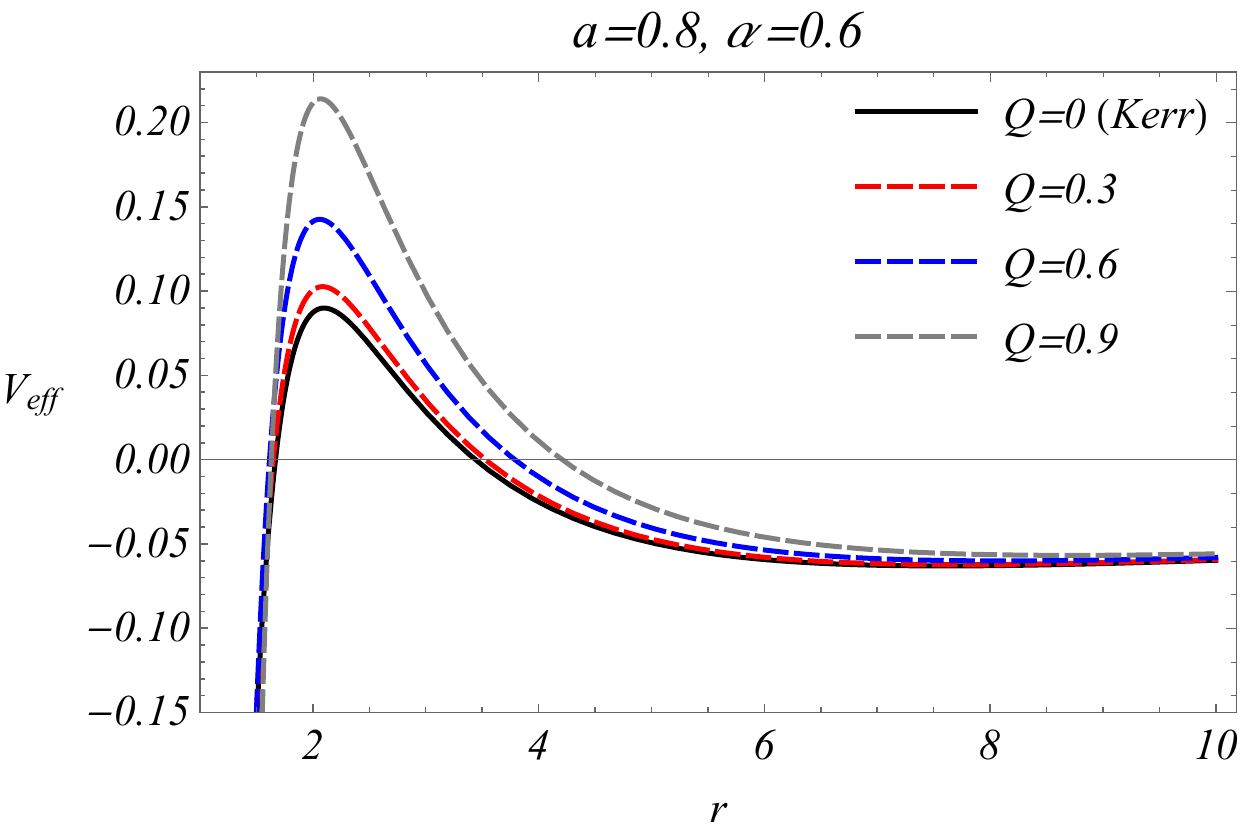}}
\subfigure{\includegraphics[width=0.4\textwidth]{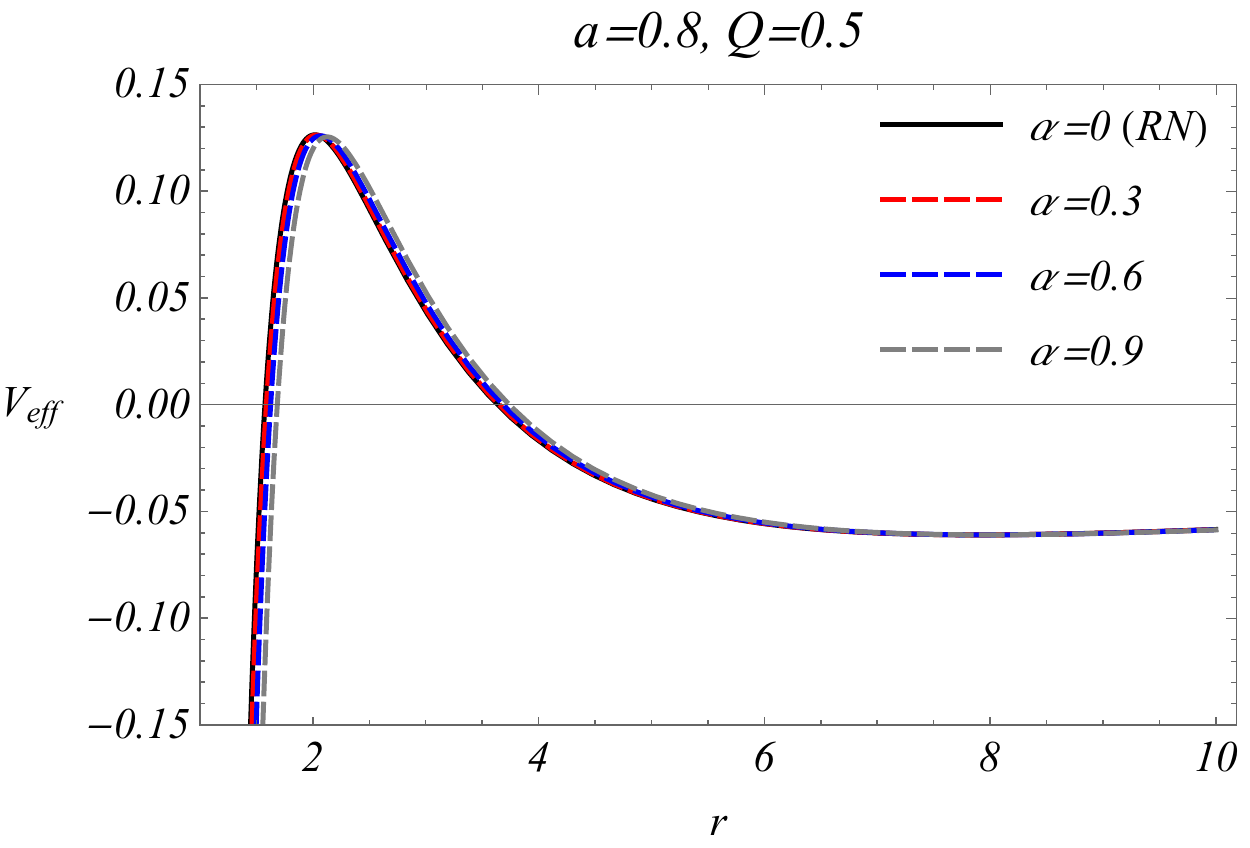}}
\caption{Behavior of stable and unstable orbits, and effective potential for various values of $a$, $\alpha$ and $Q$ with $M=1$. \label{f3}}
\end{figure*}

Massive particles in the vicinity of the rotating black hole in EMD gravity follows the timelike geodesics. In a critical region near the black hole, these particles are trapped into circular orbits governed by effective potential. We will restrict the study in the equatorial plane $\left(\theta=\pi/2\right)$ for simplicity. The Lagrangian
\begin{equation}
\mathcal{L}=\frac{1}{2}g_{\mu\nu}\dot{x}^\mu\dot{x}^\nu \label{30}
\end{equation}
and the momenta
\begin{equation}
p_\mu=g_{\mu\nu}\dot{x}^\nu \label{31}
\end{equation}
generate two constants of motion, energy $E$ and angular momentum $L$ given as
\begin{eqnarray}
E&:=&-\frac{\partial\mathcal{L}}{\partial\dot{t}}=-g_{tt}\dot{t}-g_{t\phi}\dot{\phi}=p_t, \label{32}\\
L&:=&-\frac{\partial\mathcal{L}}{\partial\dot{\phi}}=-g_{\phi\phi}\dot{\phi}-g_{t\phi}\dot{t}=p_\phi, \label{33}
\end{eqnarray}
where, $\dot{x}=\frac{\partial x}{\partial\tau}$ and $\tau$ denotes the affine parameter. From Eqs. (\ref{32}) and (\ref{33}), the $t$ and $\phi$ components of geodesic equations become
\begin{align}
H\dot{t}&=\frac{h(r)+a^2}{\Delta(r)}\left(E\left(h(r)+a^2\right)-aL\right)-\frac{a^2E}{\csc^2\theta}+aL, \label{34}\\
H\dot{\phi}&=\frac{a}{\Delta(r)}\left(E\left(h(r)+a^2\right)-aL\right)-aE+\frac{L}{\sin^2\theta}. \label{35}
\end{align}
To determine the other two constants of motion, we incorporate the Hamilton-Jacobi equation
\begin{equation}
\frac{\partial S}{\partial\tau}=-\frac{1}{2}g^{\mu\nu}\frac{\partial S}{\partial x^\mu}\frac{\partial S}{\partial x^\nu}, \label{36}
\end{equation}
with Jacobi action
\begin{equation}
S=\frac{1}{2}m_p^2\tau-Et+L\phi+S_r(r)+S_\theta(\theta), \label{37}
\end{equation}
where, $m_p$ denotes the particle's rest mass as the third constant of motion defined as $m_p^2=-p_\mu p^\mu$. The separation of variables $r$ and $\theta$ determines the other two geodesic equations given as
\begin{eqnarray}
H\dot{r}&=&\pm\sqrt{R(r)}, \label{38}\\
H\dot{\theta}&=&\pm\sqrt{\Theta(\theta)}, \label{38a}
\end{eqnarray}
where,
\begin{align}
R(r)&=\left(aL-\left(h(r)+a^2\right)E\right)^2-\Delta(r)\left(K+m_ph(r)\right), \label{39}\\
\Theta(\theta)&=K-a^2m_p\cos^2\theta-\left(aE\sin\theta-\frac{L}{\sin\theta}\right)^2, \label{39a}
\end{align}
giving rise to the modified Carter constant $K=Z+(L-aE)^2$ \cite{PhysRev.174.1559,PhysRevD.89.124004} with $Z$ being the Carter constant treated as fourth constant of motion. For massive particles, we consider $m_p=1$. Moreover, for the equatorial orbits, $Z=0$ such that $K\geq0$. The radial equation of motion in terms of the effective potential $V_{eff}(r)$ is written as
\begin{equation}\label{40}
\dot{r}^2+2V_{eff}(r)=0,
\end{equation}
so that the explicit form of the effective potential $V_{eff}(r)$ becomes
\begin{equation}\label{41}
2h(r)^2V_{eff}(r)=-R(r).
\end{equation}
From Eqs. (\ref{38}) and (\ref{38a}), the particle motion is allowed if $R(r)\geq0$ and $\Theta(\theta)\geq0$. This implies that $V_{eff}(r)\leq0$. The bounds on the angular momentum $L$ of the particle are determined by solving the equations
\begin{equation}\label{42}
V_{eff}(r)=0=\frac{dV_{eff}(r)}{dr}.
\end{equation}
The limiting values $L^{max}$ for extremal black holes have been computed for different values of charge $Q$ and fixed $\alpha$ as given in Tab. \ref{T1}. It is observed that by increasing the value of $Q$, the extremal horizon increases by a small fraction, the extremal spin decreases, whereas, the maximum angular momentum increases. The maximum allowed values $L^{max}$ for extremal black holes for different values of charge $\alpha$ and fixed $Q$ are given in Tab. \ref{T2}. It is found that by increasing the value of $\alpha$, the extremal horizon increases, however, the extremal spin increases with a fractionally small margin and the maximum angular momentum increases. Moreover, we have calculated the effective potential for the massive particle to determine the orbital behavior of the trajectories. The circular orbits are determined by the condition (\ref{42}), and the stable orbits correspond to the condition $\partial^2V_{eff}(r)>0$, whereas, the condition $\partial^2V_{eff}(r)<0$ determines the unstable orbits. This behavior of effective potential is shown in Fig. \ref{f3} in which the local maxima and local minima correspond to the unstable and stable orbits, respectively. The two plots in the upper panel show a clear variation in the effective potential for the unstable orbits. For both plots, the effective potential increases with increase in $a$ and $Q$. However, the effective potential corresponding to the unstable orbits shown in the lower panel depicts a fractional decrease with increase in $\alpha$. Moreover the unstable orbits are shifted away from the origin with increase in $a$ and $\alpha$ as seen in upper left and the bottom plots, respectively. Whereas, the unstable orbits are shifted towards the central object with increase in $Q$ corresponding to the upper right plot. The stable orbits in the upper left plot are shifted away from the origin and are attained at increasing effective potential values with respect to increase in $a$. The same behavior is observed in the right plot with respect to increase in $Q$. In the lower plot, the stable orbits are also shifted away from the origin with respect to increase in $\alpha$, however, are attained at fractionally decreasing values of effective potentials.
\begin{table}
\centering
\begin{tabular}{|p{1.5cm}|p{1.5cm}|p{1.5cm}|p{1.5cm}|}
\hline
\multicolumn{4}{|c|}{$\alpha=0.5,M=E=1$} \\
\hline
$Q$ & $r^E_H$ & $a^E$ & $L^{max}$ \\
\hline
0.1 & 1.00125 & 0.99499 & 2.00254 \\
0.2 & 1.00504 & 0.97984 & 2.01072 \\
0.3 & 1.01145 & 0.95421 & 2.02633 \\
0.4 & 1.02064 & 0.91744 & 2.05289 \\
0.5 & 1.03287 & 0.86851 & 2.09685 \\
0.6 & 1.04853 & 0.80586 & 2.17014 \\
\hline
\end{tabular}
\caption{The maximal angular momentum of the extremal spinning black hole in EMD gravity for different values of $Q$ and fixed $\alpha$ with $E=M=1$. \label{T1}}
\end{table}
\begin{table}
\centering
\begin{tabular}{|p{1.5cm}|p{1.5cm}|p{1.5cm}|p{1.5cm}|}
\hline
\multicolumn{4}{|c|}{$Q=0.5,M=E=1$} \\
\hline
$\alpha$ & $r^E_H$ & $a^E$ & $L^{max}$ \\
\hline
0.1 & 1.00134 & 0.86612 & 2.02379 \\
0.3 & 1.01197 & 0.86694 & 2.04821 \\
0.5 & 1.03287 & 0.86851 & 2.09685 \\
0.7 & 1.06334 & 0.87073 & 2.16928 \\
0.9 & 1.10248 & 0.87347 & 2.26501 \\
\hline
\end{tabular}
\caption{The maximal angular momentum of the extremal spinning black hole in EMD gravity for different values of $\alpha$ and fixed $Q$ with $E=M=1$. \label{T2}}
\end{table}

\section{Center of Mass Energy}\label{S5}
A black hole may behave as a particle accelerator, when a particle enters its strong gravitational field. The trajectory and speed of the particle are affected by the field strength of the black hole. Therefore, when two accelerating particles collide near the event horizon, a huge amount of energy is generated. The two particle system with the center of mass also possesses a CME as a result of the collision of the two particles. Here, we consider two massive uncharged particles initially at rest and traveling in the equatorial plane from infinity towards the rotating black hole in EMD gravity. The particles possess rest masses $m_i$, energies $E_i$ and angular momenta $L_i$, where $i=1,2$. These particles are subjected to a head-on inelastic collision in the vicinity of the rotating black hole in EMD gravity carrying the 4-momentum given as
\begin{equation}\label{43}
p_i^\mu=m_iu_i^\mu,
\end{equation}
where, $u_i^\mu$ is the 4-velocity of the $ith$ particle. Hence, the total 4-momentum of the system becomes
\begin{equation}
p_T^\mu=p_{(1)}^\mu+p_{(2)}^\mu. \label{44}
\end{equation}
As we know, in the center of mass frame, the spatial components of the 4-momentum vanish, so the CME of the system becomes \cite{2015JHEP...05..147Z}
\begin{align}
\epsilon_{cm}^2&=-p_T^\mu p_{T\mu}\nonumber\\&=-\left(m_1u^\mu_{(1)}+m_2u^\mu_{(2)}\right)\left(m_1u_{(1)\mu}+m_2u_{(2)\mu}\right). \label{45}
\end{align}
Using the normalization condition $u_{(i)}^\mu u_{(i)\mu}=-1$ in the Eq. (\ref{45}), one obtains
\begin{equation}
\epsilon_{cm}^2=2m_1m_2\left(\frac{\left(m_1-m_2\right)^2}{2m_1m_2}+1-g_{\mu\nu}u_{(1)}^\mu u_{(2)}^\nu\right). \label{46}
\end{equation}
When the particles have equal masses, then we consider $m_0=m_1=m_2$ and the relation for CME becomes
\begin{equation}
\epsilon_{cm}^2=2m_0^2\left(1-g_{\mu\nu}u_{(1)}^\mu u_{(2)}^\nu\right). \label{47}
\end{equation}
The 4-velocities of the two particles can be obtained from the geodesic equations (\ref{34}), (\ref{35}), (\ref{38}) and (\ref{38a}) such that
\begin{equation}
u^\mu_i=\left(\dot{t}_i,\dot{r}_i,0,\dot{\phi}_i\right), \label{48}
\end{equation}
where,
\begin{figure*}
\centering
\subfigure{\includegraphics[width=0.4\textwidth]{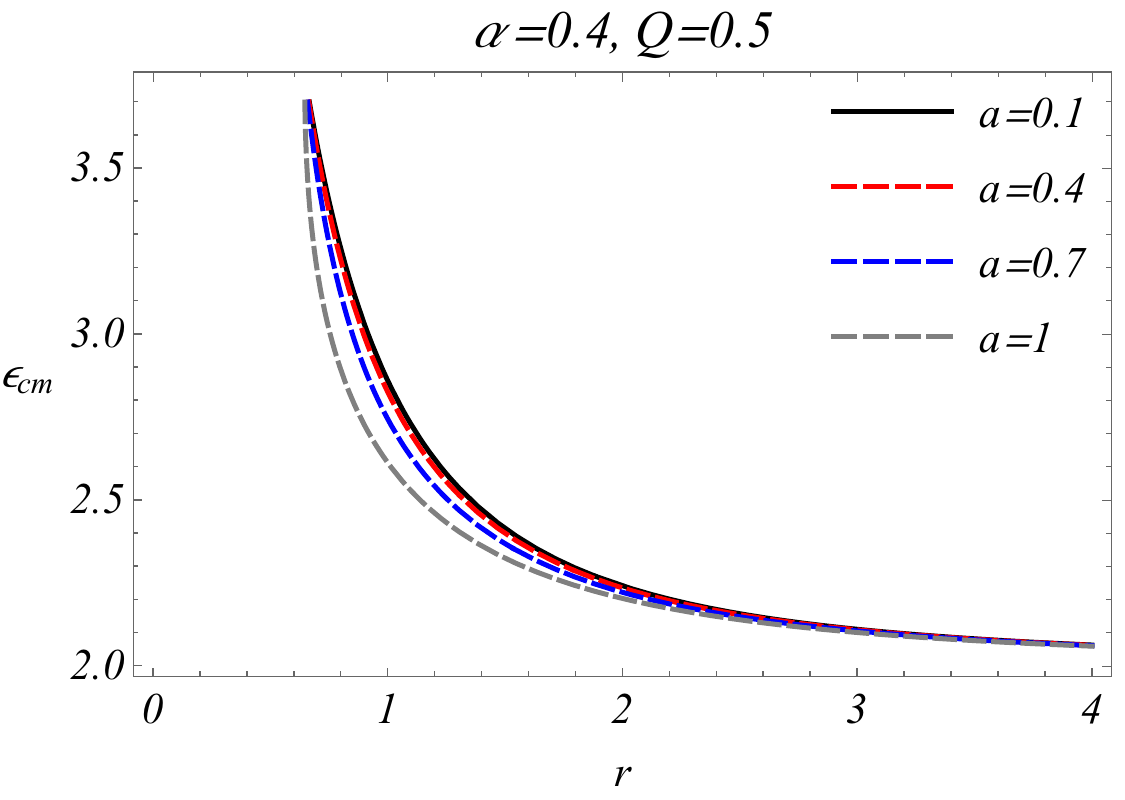}}~~~
\subfigure{\includegraphics[width=0.4\textwidth]{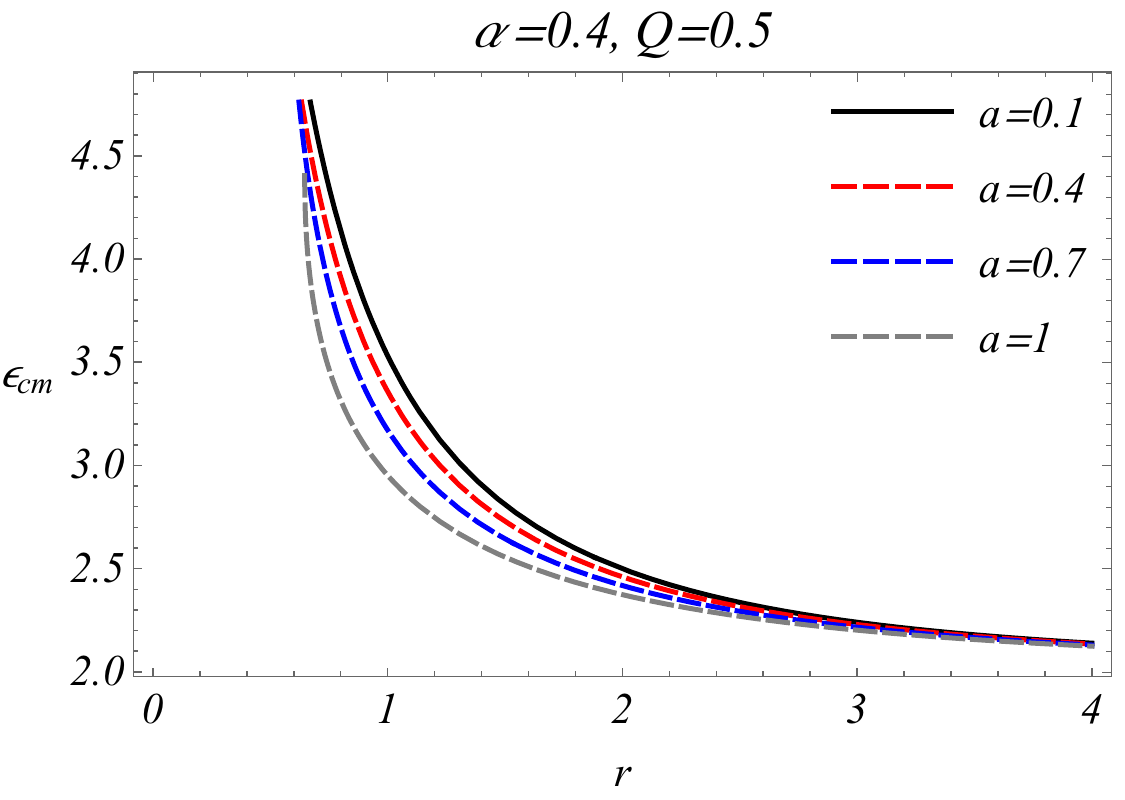}}
\subfigure{\includegraphics[width=0.4\textwidth]{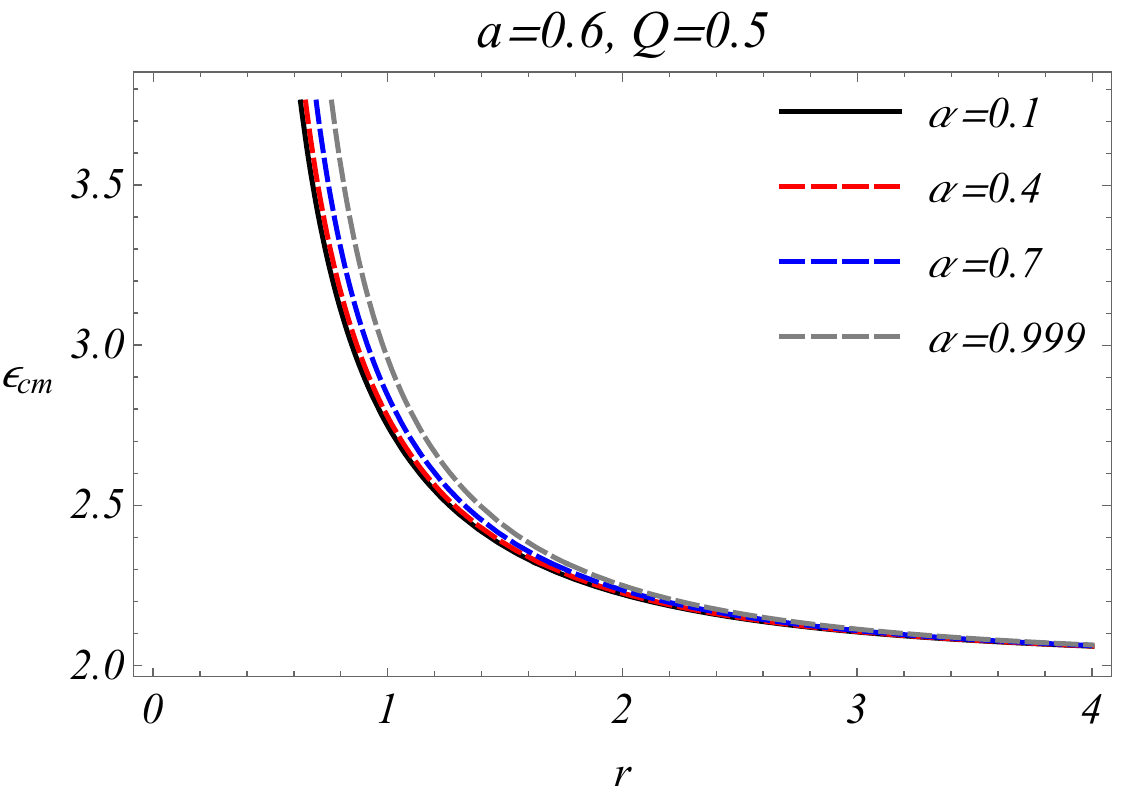}}~~~
\subfigure{\includegraphics[width=0.4\textwidth]{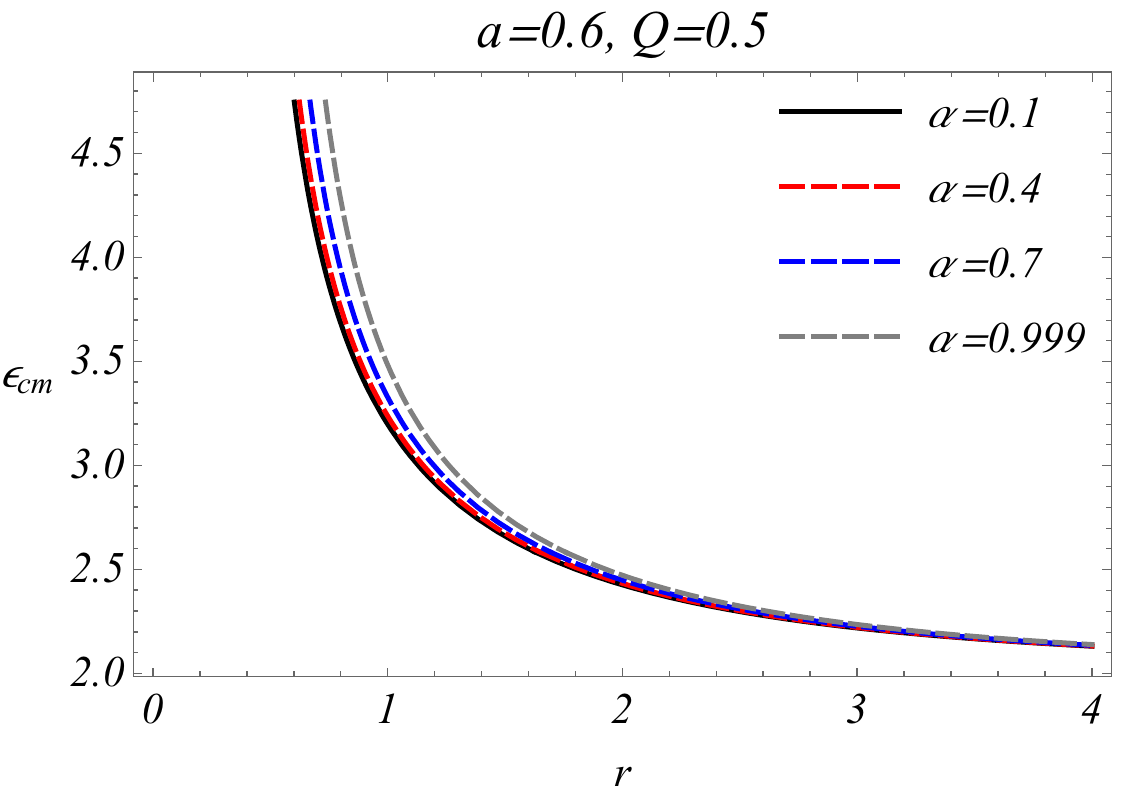}}
\subfigure{\includegraphics[width=0.4\textwidth]{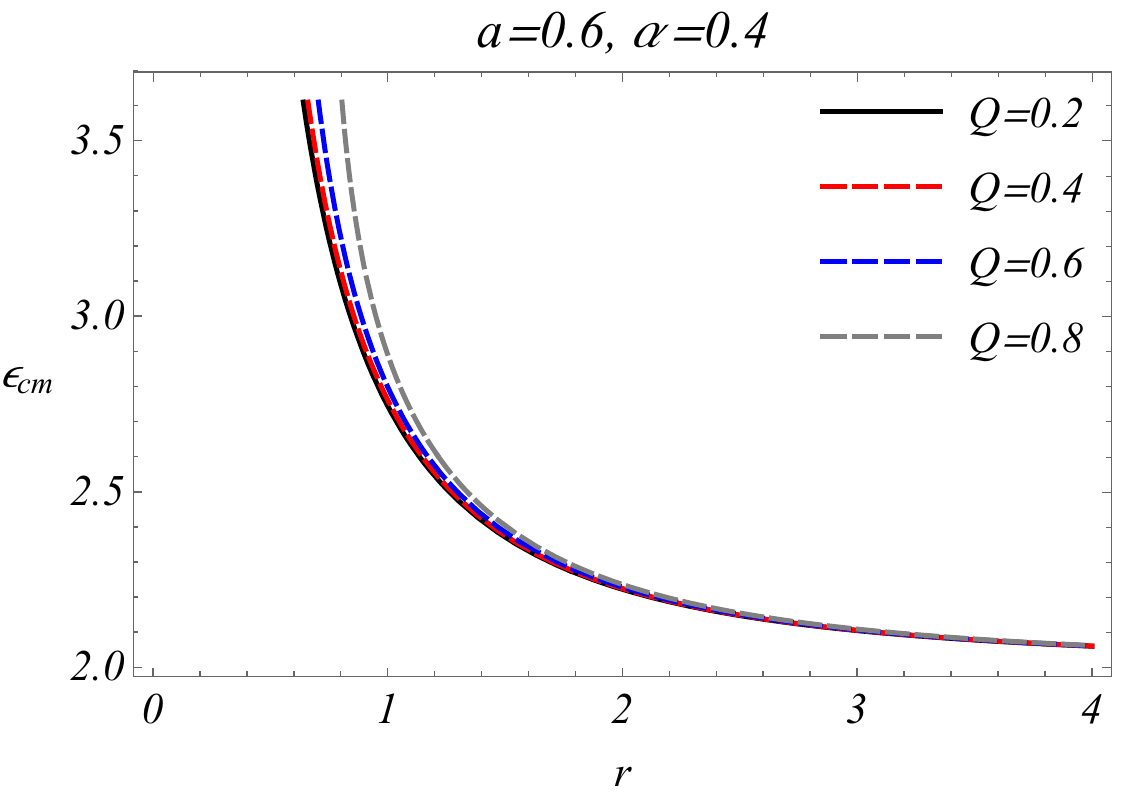}}~~~
\subfigure{\includegraphics[width=0.4\textwidth]{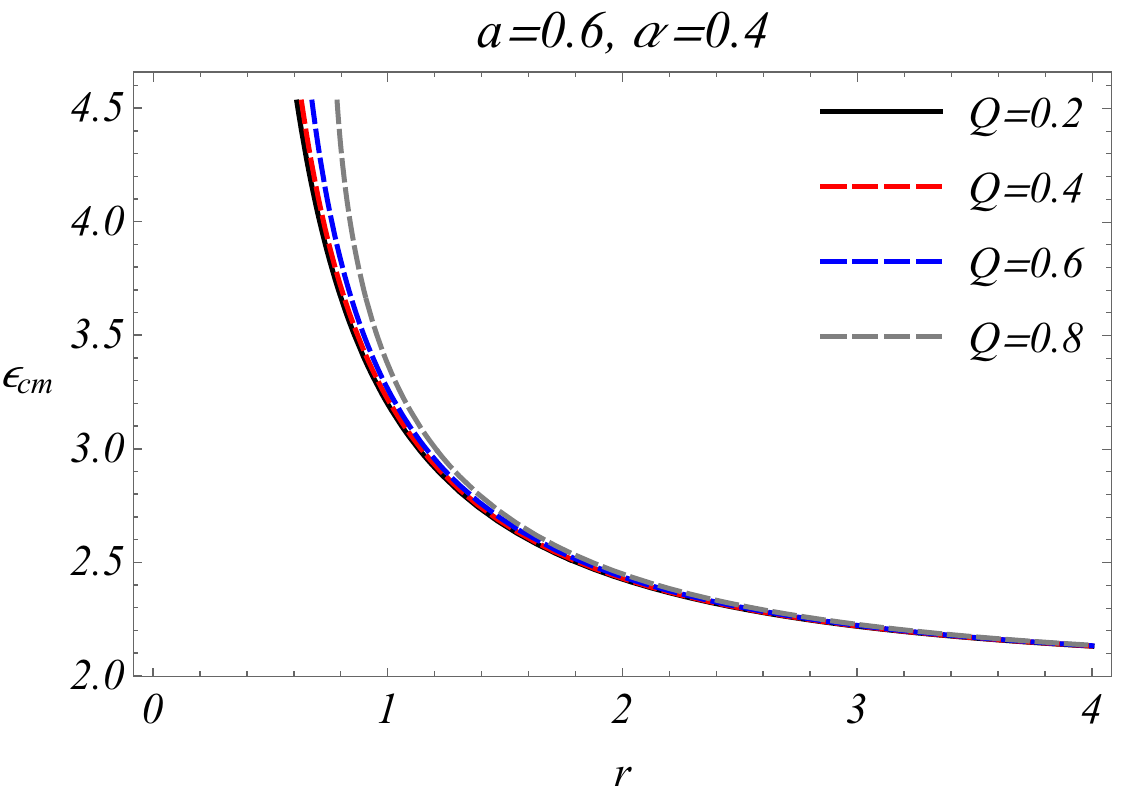}}
\caption{Plots showing the CME extracted from the collision of two massive neutral particles in the equatorial plane outside the event horizon of the rotating black hole in EMD gravity. The colliding particles have equal masses $m_0=1$ and angular momentum $L_1=1$. Moreover, the left panel corresponds to $L_2=-1$ and for the right panel $L_2=-2$. \label{f4}}
\end{figure*}
\begin{figure*}
\centering
\subfigure{\includegraphics[width=0.4\textwidth]{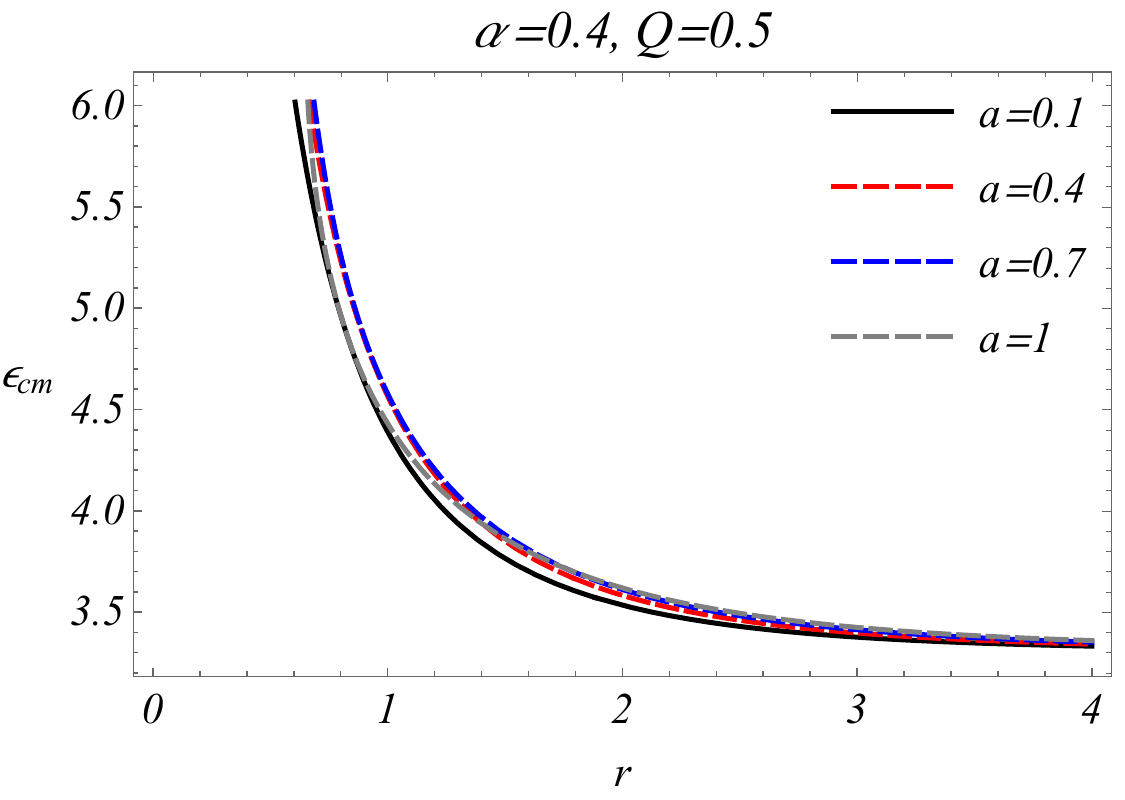}}~~~
\subfigure{\includegraphics[width=0.4\textwidth]{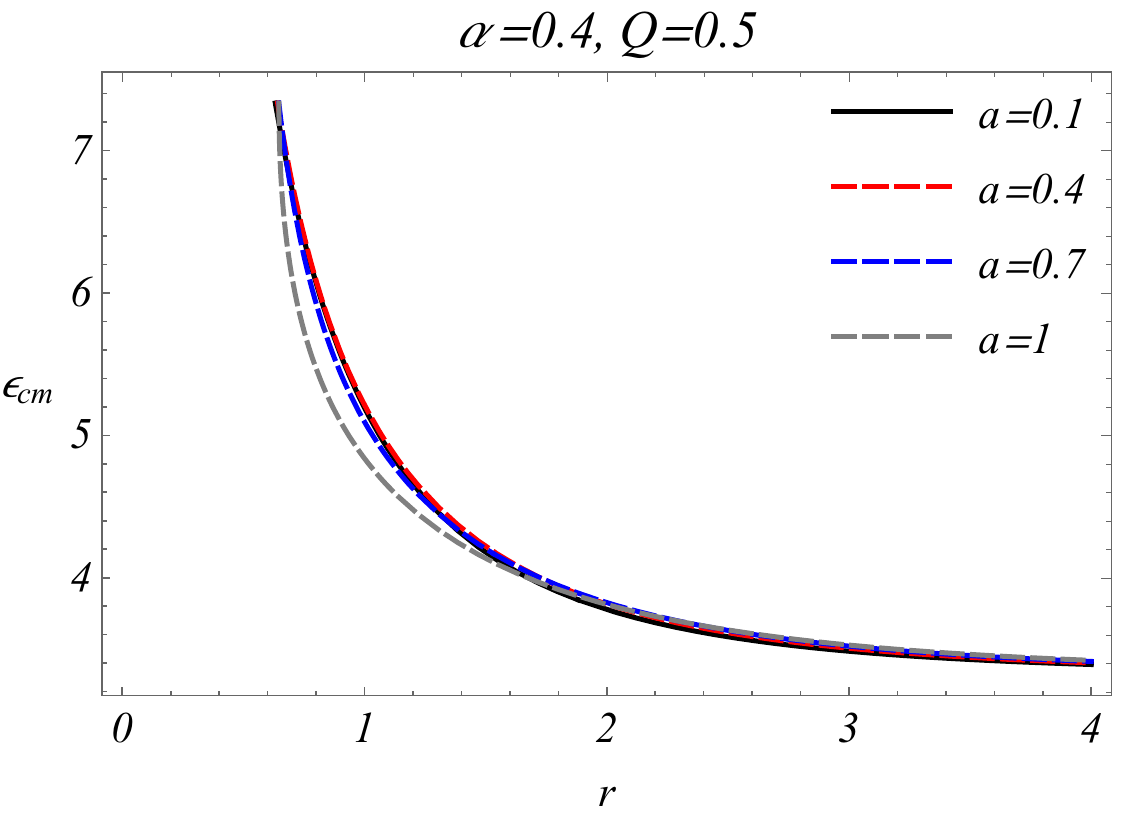}}
\subfigure{\includegraphics[width=0.4\textwidth]{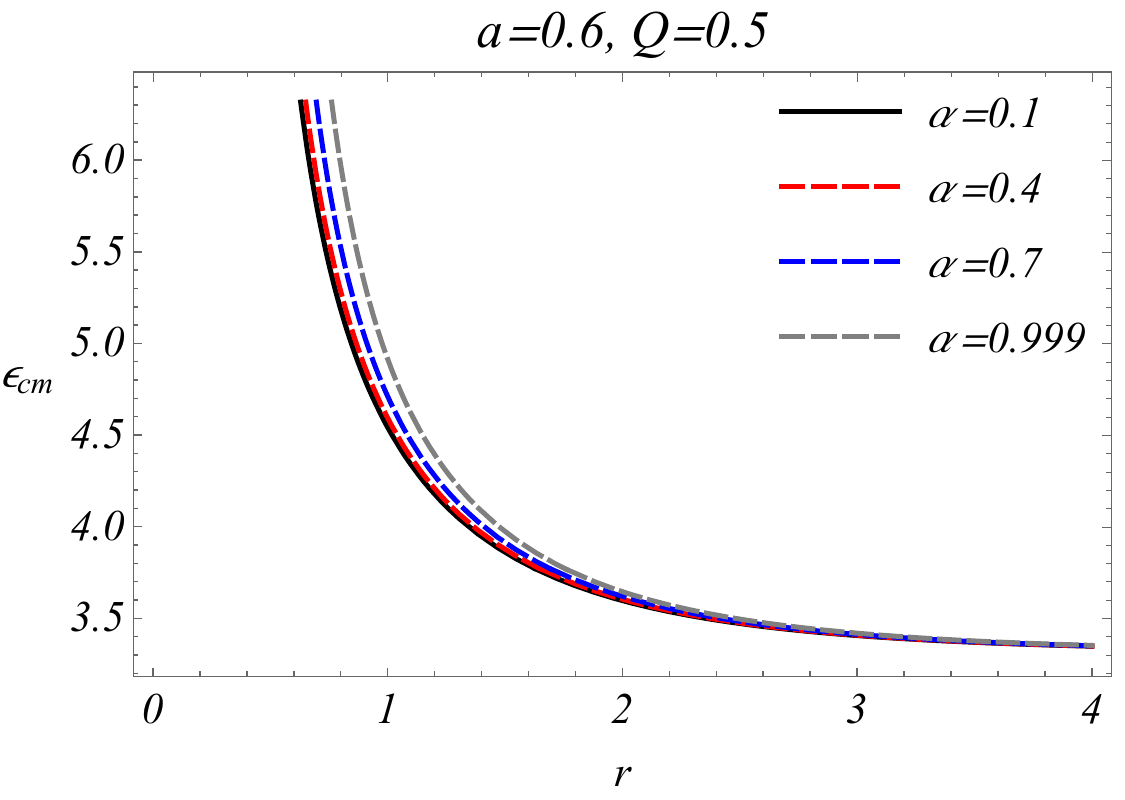}}~~~
\subfigure{\includegraphics[width=0.4\textwidth]{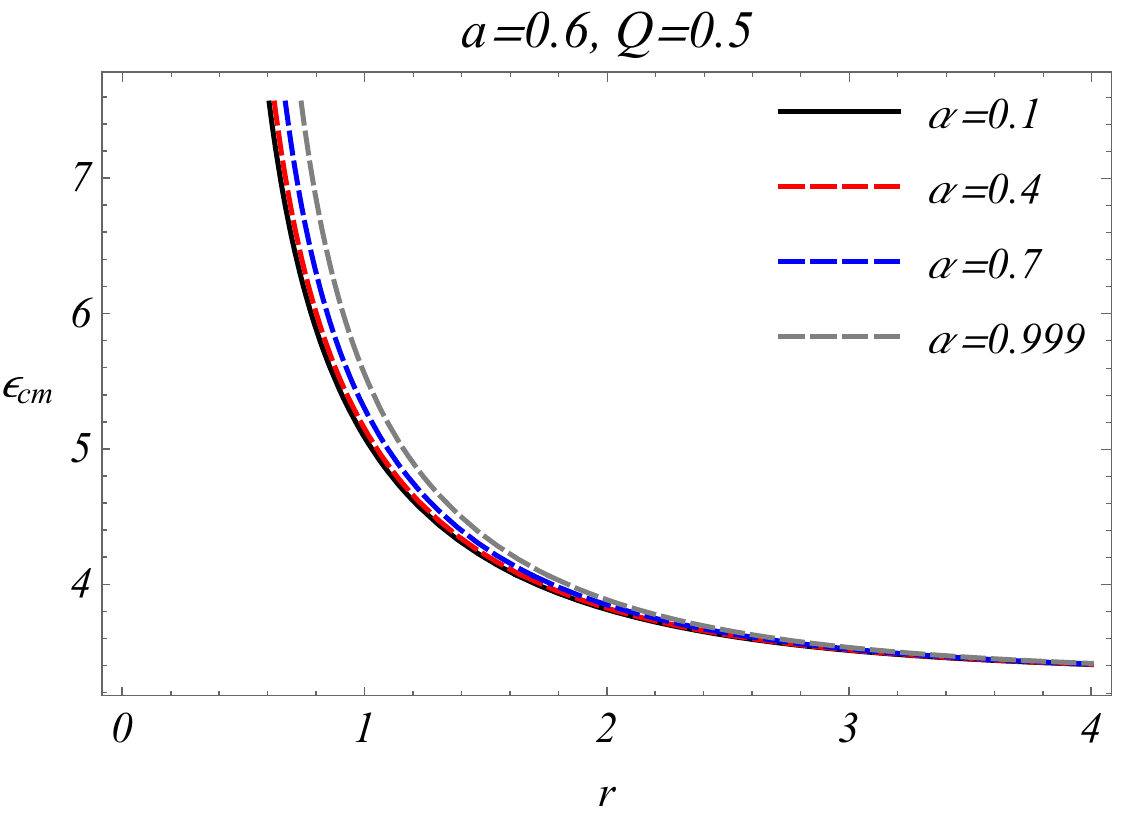}}
\subfigure{\includegraphics[width=0.4\textwidth]{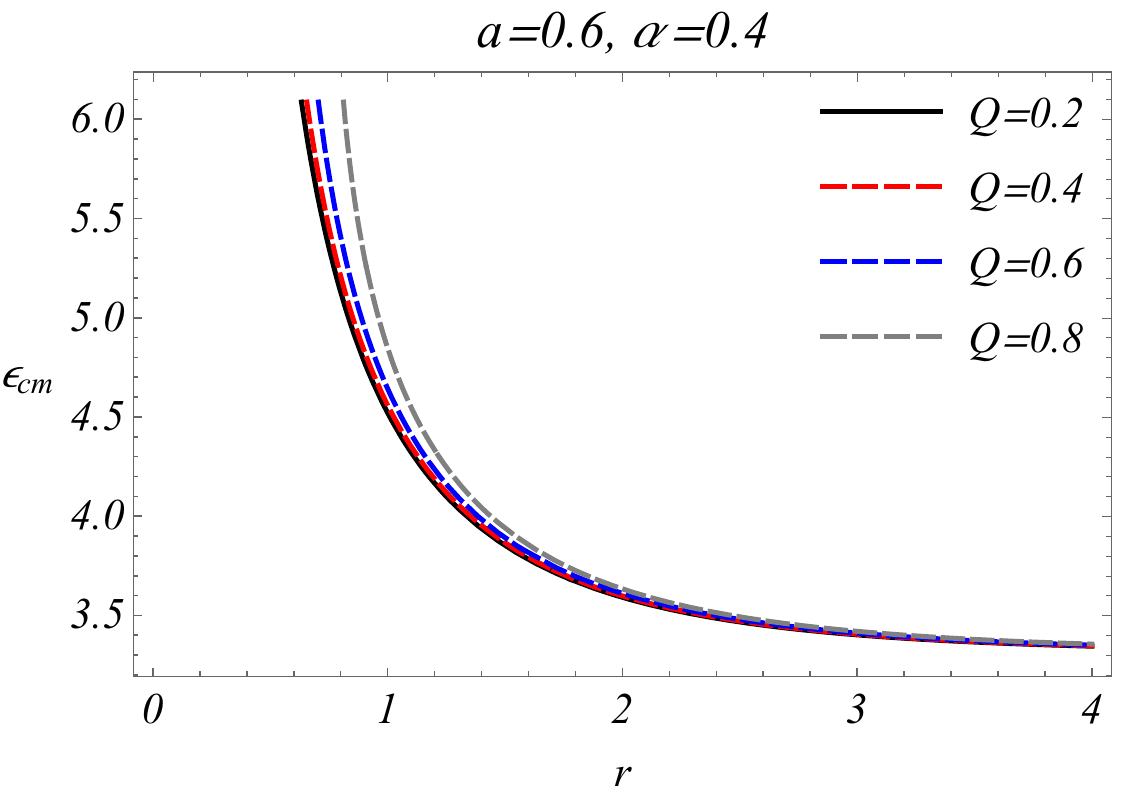}}~~~
\subfigure{\includegraphics[width=0.4\textwidth]{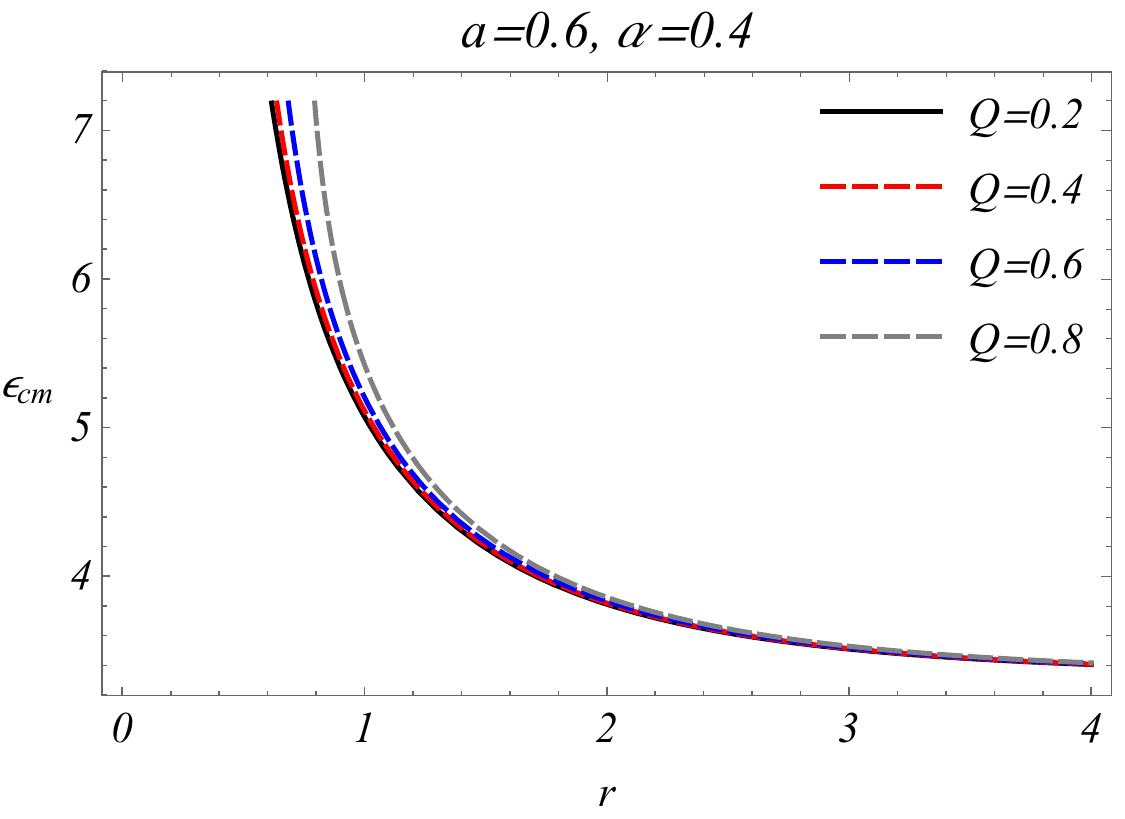}}
\caption{The behavior of CME extracted from the collision of two uncharged massive particles in the equatorial plane outside the event horizon of the rotating black hole in EMD gravity. The colliding particles have different masses $m_1=1$ and $m_2=2$, and angular momentum $L_1=1$. Moreover, the left panel corresponds to $L_2=-1$ and for the right panel $L_2=-2$. \label{f5}}
\end{figure*}
\begin{align}
\dot{t}_i&=\frac{h(r)+a^2}{H\Delta(r)}\left(E_i\left(h(r)+a^2\right)-aL_i\right)-\frac{a^2E_i-aL_i}{H}, \label{49}\\
\dot{r}_i&=\pm\frac{\sqrt{\left(aL_i-\left(h(r)+a^2\right)E_i\right)^2-\Delta(r)\left(K_i+h(r)\right)}}{H}, \label{50}\\
\dot{\phi}_i&=\frac{a}{H\Delta(r)}\left(E_i\left(h(r)+a^2\right)-aL_i\right)+\frac{L_i-aE_i}{H}. \label{51}
\end{align}
The expressions for CME become extremely complicated due to the metric tensor being complicated for the rotating metric in EMD gravity. Therefore, it is quite hard to derive expressions for near horizon collision. Hence, we calculate general behavior of the collisions in the vicinity of the event horizon. In Fig. \ref{f4}, we have presented the behavior of CME $\epsilon_{cm}$ for the collision of uncharged particles with equal masses. The plots in the left panel show the head-on collision at same speed of both particles. Whereas, the right panel corresponds to the collision at different speed of the particles. For all cases, the CME increases as the collision takes place near the black hole. In the upper panel, the curves are varied for different values of spin $a$, keeping $\alpha$ and $Q$ fixed. It shows that far from the black hole, the CME does not vary significantly with increase in $a$. However, as the collision takes place near the black hole, the variation in CME increases. With increase in the angular momentum in the panel, the variation in CME increases at a particular radial location. In the middle panel, with respect to $\alpha$, almost the same behavior of CME is observed as in the upper panel. However, when the particles have different angular momenta, the CME rises more rapidly closer to the black hole. A similar behavior of CME with respect to angular momenta and charge $Q$ is observed in the lower panel as in upper and middle panels. The Fig. \ref{f5} is identical with the Fig. \ref{f4} but for the collision of particles with different masses. For all cases, it is found that the CME is relatively higher at a particular radial location as compared to the case discussed in Fig. \ref{f4}. Moreover, the CME rises with accelerated rate as the collision approaches the black hole. Furthermore, the CME is greater at a radial location for particles with different angular momenta as compared to the corresponding cases in left panel. In the upper panel, the variation in the CME with respect to $a$ is not obvious and is diminished in comparison with the Fig. \ref{f4}. However, the variation with respect to $\alpha$ and $Q$ in middle and lower panels is same as that in Fig. \ref{f4}.

\section{Center of Mass Energy in Plasma}\label{S6}
The outer space comprise various fluids in excited states generally termed as plasma. In particular, due to sufficiently high gravity, the matter may exist around a black hole creating a high temperature due to interactions. This may lead to the existence of plasma around a black hole. Therefore, we intend to study the collision of particles and energy extraction from their center of mass in the presence of plasma. In Sec. \ref{S5}, the BSW mechanism was incorporated for massive particles in non-plasma medium. However, we need some assumptions, as the BSW mechanism does not work in general in case of plasma medium. If we consider particles with non-zero rest masses in vacuum, the condition for their propagation in plasma cannot be determined by the refractive index relation that works only for massless particles, specifically for photons. On the other hand, if we consider massless particles to ensure the propagation condition but one may assert that the BSW mechanism fails for the zero rest mass of the particles. However, this is not true as the photons and other massless particles become timelike with non-zero rest masses in plasma. In this way one can determine the condition for particle propagation so that the BSW mechanism can also be applied. First, we need to determine the refractive index of the medium that provides the propagation condition of the particle through plasma. We consider a free de Broglie particle with rest mass $m_r$ moving in a field such as a dispersive medium of potential $V=constant$ carrying a matter wave. The energy $E$ of the particle carrying momentum $p$ is given as
\begin{equation}
E^2=(pc)^2+(m_rc^2)^2, \label{55}
\end{equation}
with $c$ being the speed of light in vacuum. The energy and momentum for the matter wave are
\begin{equation}
E=\hbar\omega, \quad \textbf{p}=\hbar\textbf{k}, \label{56}
\end{equation}
where, $\omega$ is the angular frequency of the wave, $\hbar$ is the Planck's constant and $\textbf{k}$ is the wavevector with magnitude denoted as $k$ termed as wave number. We have considered the relation $E=\hbar\omega$ which generally describes the energy of a photon in vacuum. However, this is not true in general, that is, the relation $E=\hbar\omega$ holds for other quantum particles as well. It can be deduced from the fact that for the time evolution of a quantum state, the Schrodinger's equation is written as
\begin{equation}
i\hbar\frac{d\ket{\Psi}}{dt}=H\ket{\Psi}. \label{57}
\end{equation}
Since, we are dealing with a free particle, therefore, the state $\ket{\Psi}$ has energy states as plane waves for which the Eq. (\ref{57}) takes the form
\begin{equation}
i\hbar\frac{d\ket{\Psi}}{dt}=E\ket{\Psi}. \label{58}
\end{equation}
Further, if the state is expressed in the form $e^{-i\omega t}$, then we get $E=\hbar\omega$. Now, using Eq. (\ref{56}) in (\ref{55}) gives the dispersion relation
\begin{equation}
\frac{\omega^2}{k^2}=\frac{c^2}{1-\frac{m_r^2c^4}{\hbar^2\omega^2}}. \label{59}
\end{equation}
By comparing the Eq. (\ref{59}) with the general dispersion relation $\frac{\omega^2}{k^2}=\frac{c^2}{n^2}$, we get
\begin{equation}
n^2=1-\frac{m_r^2c^4}{\hbar^2\omega^2}. \label{60}
\end{equation}
In natural units, $c=1=\hbar$, the relation (\ref{60}) for optical refractive index for a de Broglie particle reduces to
\begin{equation}
n^2=1-\frac{m_r^2}{\omega^2}. \label{61}
\end{equation}
The Eq. (\ref{61}) is dimensionally consistent, since it follows from Eq. (\ref{60}) in which the last term is dimensionless. Since, a plasma is a dispersive and inhomogeneous medium, therefore, the photons also may not follow the null geodesics. This is why the photons in plasma are also described as particles with variable mass and their rest masses may not be zero. Hence, the above relations for refractive index can be considered for both massive and massless particles moving in a plasma medium. However, we consider Eq. (\ref{61}) with $r$ and $\theta$ dependence for the collision of photons in the plasma with non-zero rest masses, that is,
\begin{equation}
n^2(r,\theta)=1-\frac{m_r^2(r,\theta)}{\omega^2(r,\theta)}. \label{63}
\end{equation}
The photons in the plasma may not follow the null geodesics because of non-zero rest mass, but may follow the timelike geodesics. However, a suitable re-scaling of the spacetime metric can lead to null geodesics of the photons in the plasma. Since, we want to treat the colliding photons as massive particles, therefore we are not interested in the re-scaling of the black hole metric. The photon propagation in plasma is possible only if the energy of photons is greater than the energy of electrons in plasma. The observer is kept static with no spatial motion, the 4-velocity then becomes $U^\mu(r,\theta)=U^t(r,\theta)=\left(-g_{tt}(r,\theta)\right)^{-1/2}$. Therefore, by virtue of the expression $E=\hbar\omega$, the frequency of the particle measured by the static observer is $\omega(r,\theta)=-p_\mu U^\mu(r,\theta)=-p_t U^t(r,\theta)$. As we know that $p_t$ is a constant of motion, therefore, considering $p_t=-\omega_0$ gives the physical meaning of $\omega_0$. It is the frequency of the particle reaching the infinity as measured by an observer on $t$-line at infinity. Therefore, we obtain
\begin{equation}
\omega(r,\theta)=\frac{\omega_0}{\sqrt{-g_{tt}(r,\theta)}}. \label{64}
\end{equation}
The Hamilton-Jacobi equation can be written as
\begin{equation}
2\partial_{\tau}\mathcal{S_J}+g^{\mu\nu}\partial_{x^\mu}\mathcal{S_J}\partial_{x^\nu}\mathcal{S_J}+m_r^2(r,\theta)=0, \label{67}
\end{equation}
where, $\mathcal{S_J}$ is the Jacobi action given as
\begin{figure*}
\centering
\includegraphics[width=0.5\textwidth]{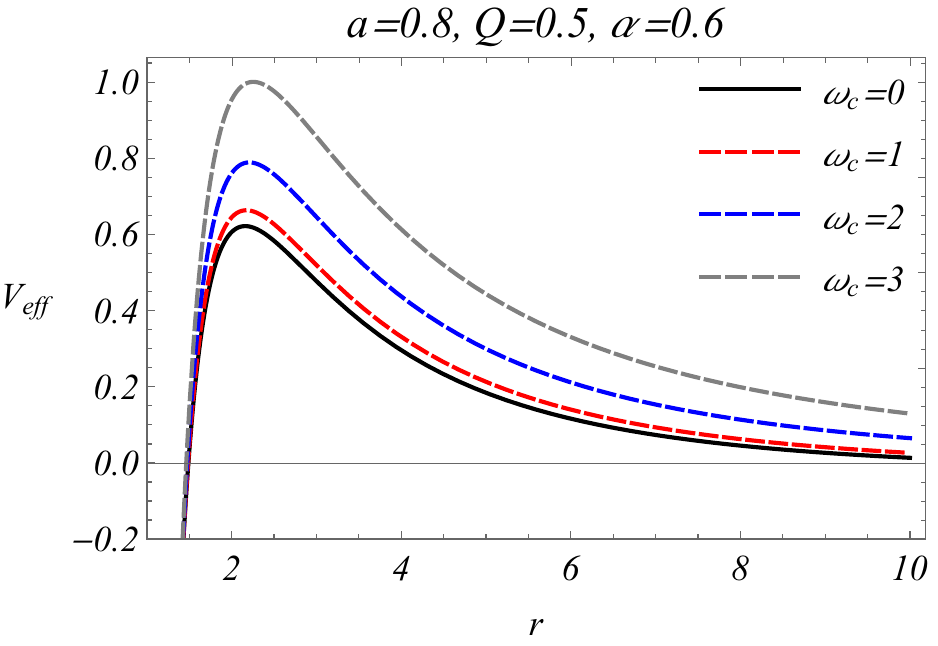}
\caption{Behavior of unstable orbits and effective potential for various values of $\omega_c$, keeping $a$, $\alpha$ and $Q$ fixed with $M=1$. \label{f6}}
\end{figure*}
\begin{equation}
\mathcal{S_J}=\frac{1}{2}m_p\tau-Et+L\phi+\mathcal{A}_r(r)+\mathcal{A}_\theta(\theta), \label{68}
\end{equation}
such that the functions $\mathcal{A}_r(r)$ and $\mathcal{A}_\theta(\theta)$ are arbitrary and unknown. Note that the particle motion undergoes within plasma where photons and other massless particles are massive, therefore the Jacobi action comprises the mass term. Generally, the last term in the Hamilton-Jacobi equation is electron frequency term which can be regarded as the effective mass of the particle under the mass-energy and Planck's relations. On further simplification, the Hamilton-Jacobi equation gives rise to a mixed term in $r$ and $\theta$ coordinates which is in general not separable, that is,
\begin{align}
&\Delta(r)\left(\partial_r\mathcal{A}_r(r)\right)^2+\left(\partial_\theta\mathcal{A}_\theta(\theta)\right)^2-\frac{\left((h(r)+a^2)\omega_0-aL\right)^2}{\Delta(r)}\nonumber\\&+(L-a\omega_0)^2+\left(L^2\csc^2\theta-a^2\omega_0^2\right)\cos^2\theta+Hm_r^2(r,\theta)\nonumber\\&+Hm_p=0. \label{69}
\end{align}
Since, $m_r^2(r,\theta)$ is mixed in $r$ and $\theta$ in general, so the Eq. (\ref{69}) is not separable. Therefore, we assume
\begin{equation}
Hm_r^2(r,\theta)=f_r(r)+f_\theta(\theta), \label{70}
\end{equation}
so that the Eq. (\ref{69}) becomes separable. Hence, we get
\begin{align}
H\dot{t}&=\frac{h(r)+a^2}{\Delta(r)}\left(\omega_0\left(h(r)+a^2\right)-aL\right)-\frac{a^2\omega_0}{\csc^2\theta}+aL, \label{65}\\
H\dot{r}&=\sqrt{\mathcal{R}(r)}, \label{71}\\
H\dot{\theta}&=\sqrt{\Theta(\theta)}, \label{72}\\
H\dot{\phi}&=\frac{a}{\Delta(r)}\left(\omega_0\left(h(r)+a^2\right)-aL\right)-a\omega_0+L\csc^2\theta, \label{66}
\end{align}
where,
\begin{align}
\mathcal{R}(r)&=\left(\left(h(r)+a^2\right)\omega_0-aL\right)^2\nonumber\\&-\Delta(r)\left(\mathcal{Z}+\left(L-a\omega_0\right)^2+h(r)m_p+f_r(r)\right), \label{73}\\
\Theta(\theta)&=\mathcal{Z}+\left(a^2\omega_0^2-L^2\csc^2\theta\right)\cos^2\theta-f_\theta(\theta)\nonumber\\&-m_pa^2\cos^2\theta. \label{74}
\end{align}
The Eqs. (\ref{65}) and (\ref{66}) further reduces to Eqs. (\ref{34}) and (\ref{35}) in essence of the Planck's relation for $\hbar=1$. Here, we have determined four constants of motion $\omega_0=E$, angular momentum $L$, mass of particle $m_p$ and the Carter constant $\mathcal{Z}$ such that the system of geodesic equations is completely integrable. In non-plasma media, the motion of photon corresponds to $m_p=0$. However, in plasma, a photon may not have zero mass and thus the geodesics behave as timelike trajectories. Therefore, we consider $m_p=1$ for simplicity. By considering the radial geodesic equation together with Eq. (\ref{40}), the effective potential then reads
\begin{figure*}
\centering
\subfigure{\includegraphics[width=0.4\textwidth]{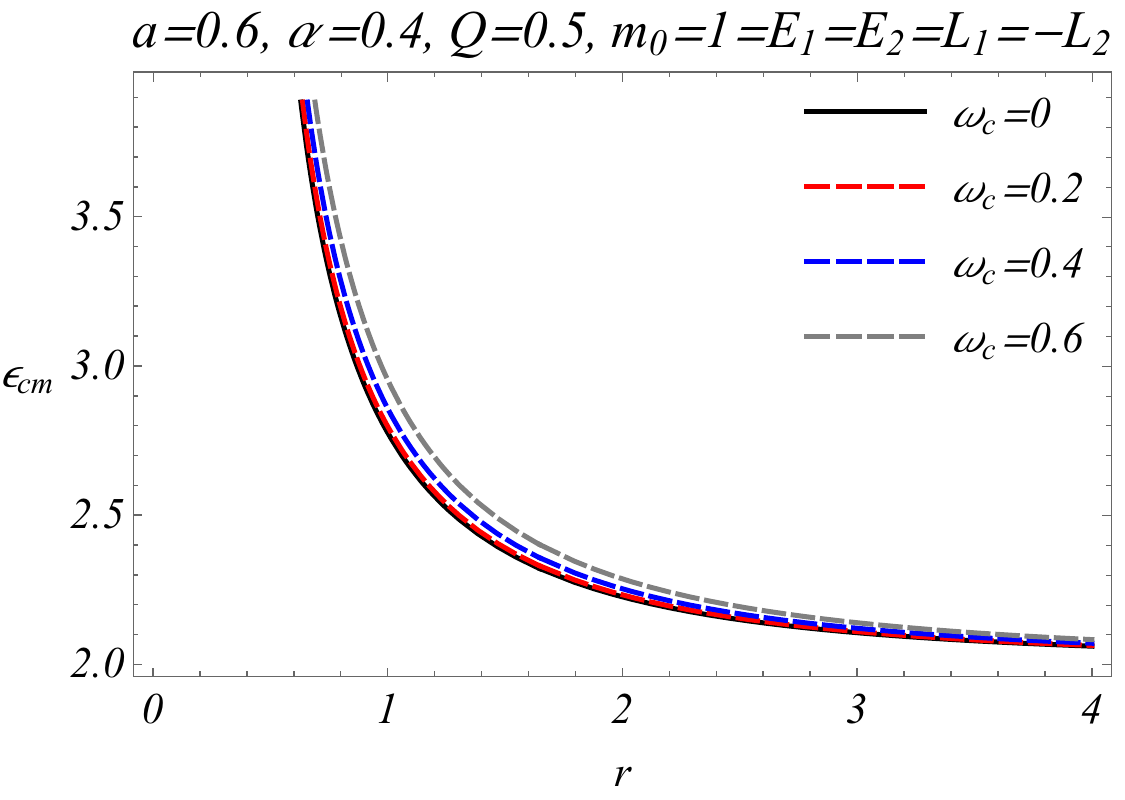}}~~~
\subfigure{\includegraphics[width=0.4\textwidth]{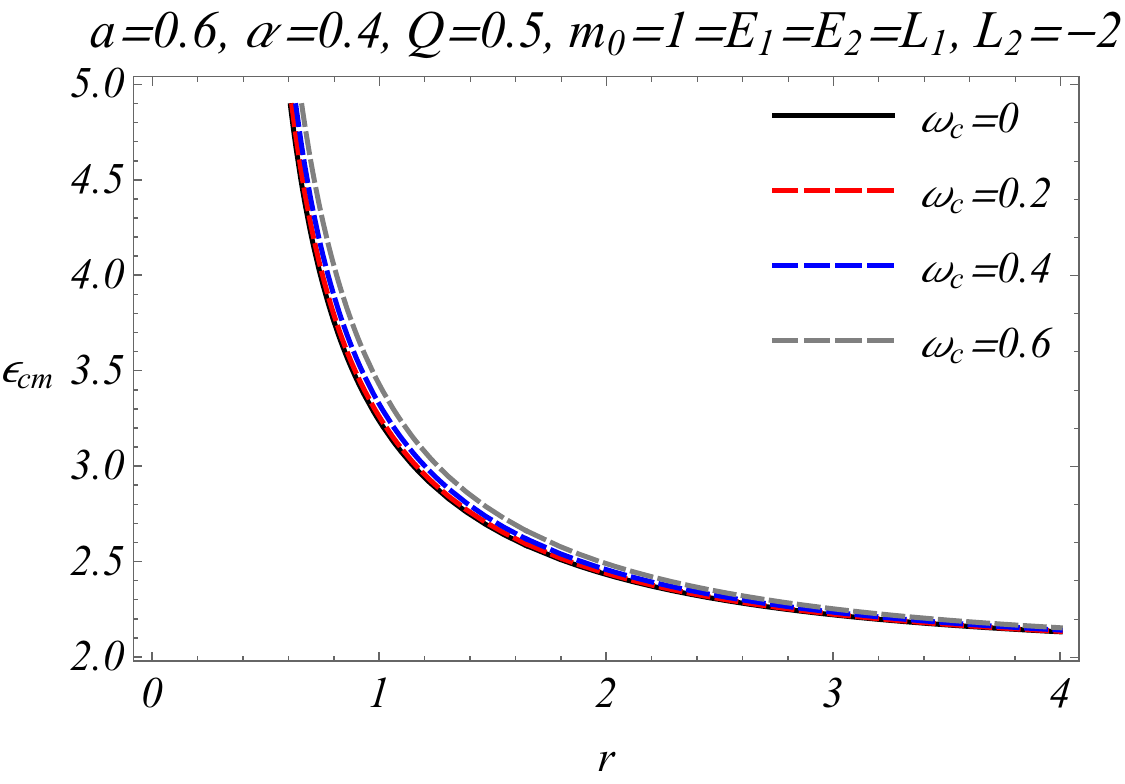}}
\subfigure{\includegraphics[width=0.4\textwidth]{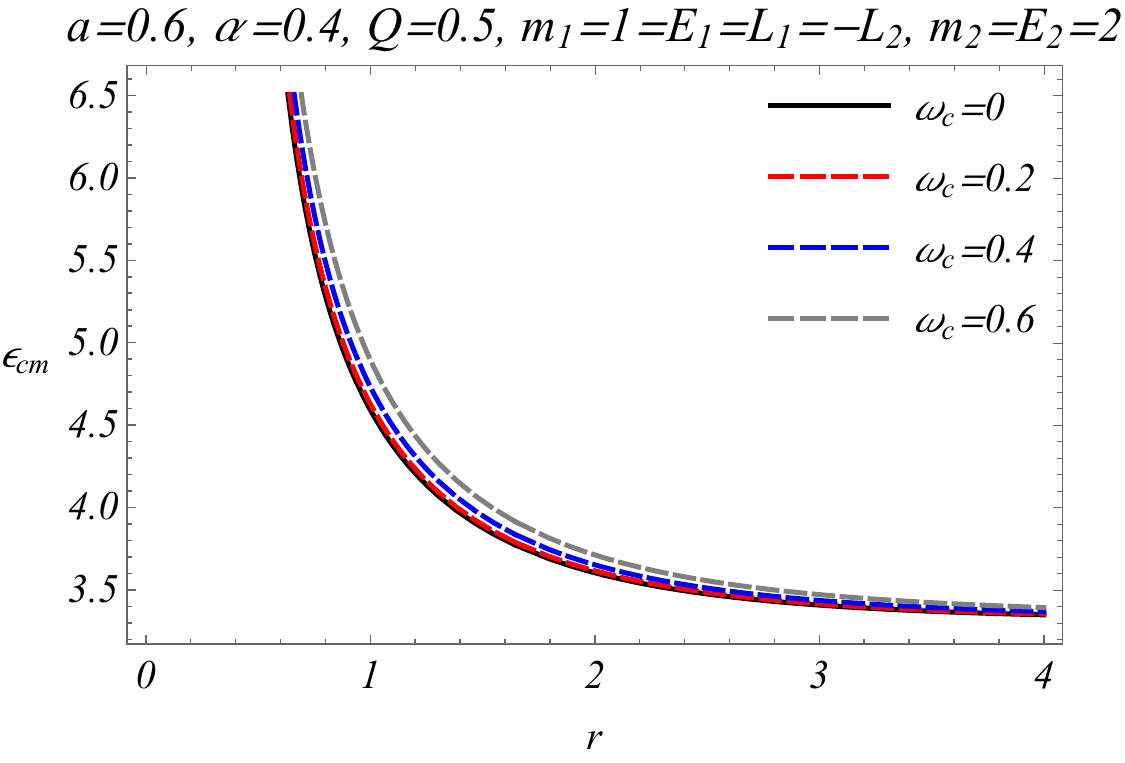}}~~~
\subfigure{\includegraphics[width=0.4\textwidth]{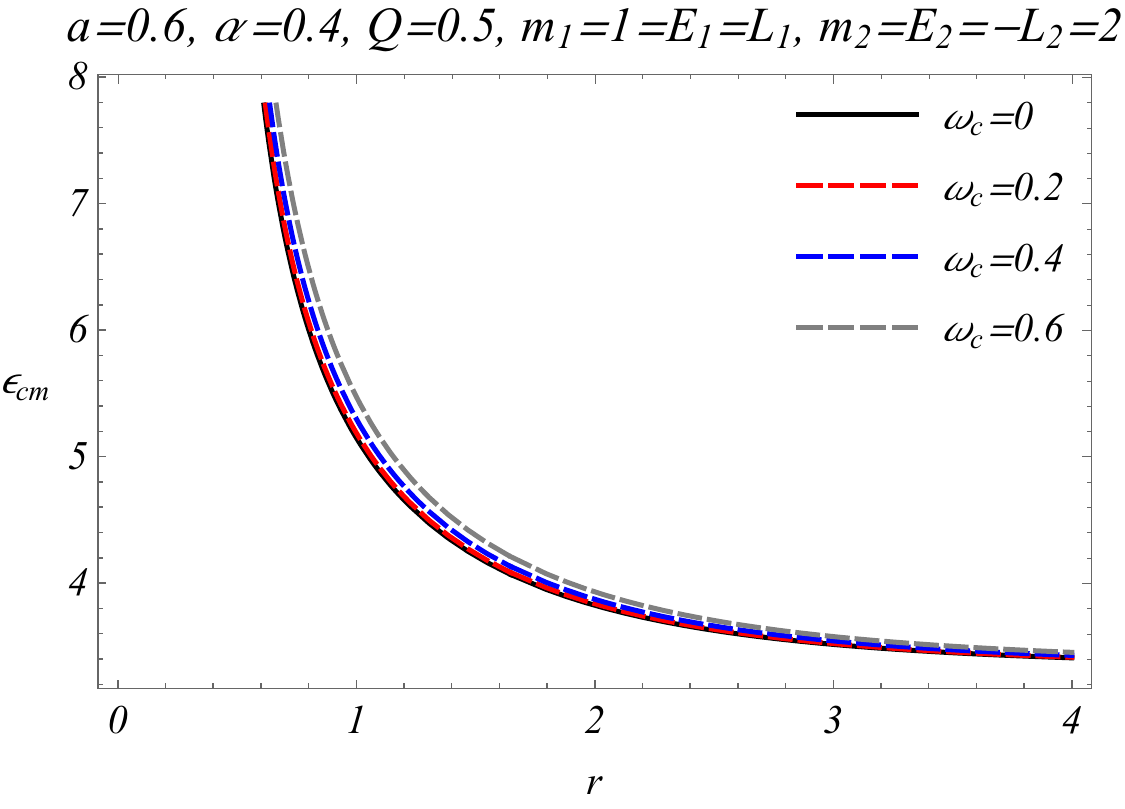}}
\caption{The plots showing the effect of plasma parameter $\omega_c$ on the behavior of CME extracted from the collision of two uncharged massive particles in the equatorial plane outside the event horizon of the rotating black hole in EMD gravity immersed in plasma. \label{f7}}
\end{figure*}
\begin{align}
V_{eff}(r)&=\frac{1}{2h(r)^2}\Big[\Delta(r)\left(\mathcal{Z}+\left(L-a\omega_0\right)^2+h(r)+f_r(r)\right)\nonumber\\&-\left(\left(h(r)+a^2\right)\omega_0-aL\right)^2\Big]. \label{75}
\end{align}
To study the effect of plasma medium on the behavior of effective potential, and the stable and unstable photon orbits, we consider $f_r(r)=\omega_c^2\sqrt{M^3r}$ \cite{PhysRevD.95.104003}. In Fig. \ref{f6}, we have shown the behavior of effective potential and the unstable orbits. The plot shows that with increase in the value of plasma parameter $\omega_c$, the unstable orbits are shifted away from the central object and are attained at relatively increasing values of effective potential for certain values of $a$, $\alpha$ and $Q$. Unlike the non-plasma case, the stable orbits are located at relatively larger values of $r$. Therefore, the graphical sketch of stable orbits is not shown. However, we have obtained the numerical values of $r$ corresponding to the stable orbits and the variation is observed for the same parametric values as in the Fig. \ref{f6}. It is found that with increase in $\omega_c$, the stable orbits are also shifted away from the origin and are attained at increasing values of effective potential.

Now, we study the collision of particles and energy extraction in the plasma medium in the vicinity of the black hole. We consider two photons initially at infinity outside the gravitational field of the black hole traveling towards the black hole with energies $E_i$, angular momenta $L_i$ and masses $m_i$, with $i=1,2$. As mentioned earlier that the photons being massless particles in vacuum posses a non-zero mass in plasma medium. Therefore, we may assume $m_i\neq0$, however, the angular momentum $L$ does not affect the intrinsic spin of the particles. Moreover, the charge of photons is also considered to be zero, such that the photons do not interact with the ions and charges in plasma medium. Note that apart from the collision of photons, one may also think of considering gluons, Z and Higgs bosons, and all three generations of neutrinos, all being uncharged particles. However, the gluons are not free particles and thus they cannot be considered for the collisions. Moreover, the lifetime of Z and Higgs bosons is too short for which it is certainly unrealistic from the collision point of view. Therefore, only neutrinos can be considered apart from photons. The CME of the system is then given by the Eqs. (\ref{46}) and (\ref{47}), whereas, the $t$ and $\phi$ components of the 4-velocity vector (\ref{48}) are same as in Eqs. (\ref{49}) and (\ref{51}). However, the radial component becomes
\begin{align}
\dot{r}_i&=\frac{1}{H}\Big[\left(aL_i-\left(h(r)+a^2\right)E_i\right)^2\nonumber\\&-\Delta(r)\left(\left(L_i-aE_i\right)^2+h(r)+f_r(r)\right)\Big]^{\frac{1}{2}}. \label{76}
\end{align}
In the vicinity of rotating EMD black hole immersed in non-homogeneous plasma, the CME extracted from the collision of massive uncharged particles is plotted in Fig. \ref{f7}. For all cases, the CME decreases as the collision takes place away farther from the black hole and approaches a constant value. This result is same as that for non-plasma medium. In all plots, the variation in CME with respect to $\omega_c$ is almost same. However, in the upper panel, when the angular momenta of the second particle is increased, the CME increases more rapidly near the black hole. The same behavior is observed in the lower panel when the particle masses are different.

\section{Concluding Remarks}\label{S7}
Studying black holes as particle accelerators has been an intriguing subject over the past few years, especially since the introduction of the BSW mechanism \cite{PhysRevLett.103.111102}. This area of research could prove valuable in high-energy astrophysics. Motivated by this, we explored the rotating EMD black hole as a particle accelerator and calculated the CME both with and without a plasma medium. Initially, we provided a brief discussion of the static metric, followed by an analysis of the horizon and ergosphere for the rotating metric.

The geodesics are then analyzed to determine the orbital properties of the trajectories and to derive the relations for 4-velocities. In the non-plasma case, all parameters have a minor effect on the size of both stable and unstable orbits. However, spin and charge have a notable impact on the effective potential of the unstable orbits. In all scenarios, the stable and unstable orbits are displaced away from the black hole, except for the unstable orbits with respect to $\alpha$, where they move toward the black hole. In the case of particle collisions with equal masses, it is observed that increasing the angular momentum of the second particle leads to a rise in the CME near the black hole. Furthermore, this increase in $L_2$ also affects the spin, as the variation in CME near the black hole becomes more pronounced with respect to $a$. When the particles have different masses, a similar behavior is seen with an increase in $L_2$. However, this adversely affects the spin's influence on the variation in CME.

Under certain assumptions, we developed a framework using the BSW mechanism to examine the influence of the plasma medium on the CME. To determine the propagation condition for massive particles, we treated photon collisions as massive particle collisions in plasma, without rescaling the metric. Both the unstable and stable orbits are displaced away from the black hole as a result of the plasma parameter. However, the influence of $\omega_c$ on the unstable orbits is weak, while it has a more substantial effect on the stable orbits and the effective potential. Additionally, $\omega_c$ exerts a weak and nearly uniform impact on the CME across all cases, indicating the expected dissipation of energy within the plasma. The plasma has a weak influence on dilaton field via coupling parameter $\alpha$.

In this study, the CME near the black hole grows significantly. This work could prove valuable in high-energy astrophysics. As a future direction, one could investigate more general plasma distributions to broaden the applicability of these results. One may particularly be interested in modifying the BSW scheme to eliminate the need for assumptions when studying particle collisions in a plasma medium, potentially incorporating dissipation factors to account for energy loss. We know about the evidence of magnetic field and hence the existence of plasma in the surrounding of M87*. Our analysis could be extended for the collision of particles in external magnetic field, and the magnetized or spinning charged particles in a magnetic field, all immersed in plasma to probe its influence on such particles and external magnetic fields. This study would lead to a comparison with EHT data for M87* that may help in some constraining studies.

\section*{Acknowledgments}
The work of M. Zubair has been partially supported by the National Natural Science
Foundation of China under project No. 11988101. He is grateful to compact object and
diffused medium Research Group at NAOC led by Prof. JinLin Han for excellent hospitality and friendly environment.

\section*{Conflict of Interest}

The authors declare no conflict of interest.

\section*{Data Availability Statement}

Data sharing is not applicable to this article as no new data were created or analyzed in this study.

\section*{Keywords}

Einstein-Maxwell-Dilaton black hole; Center of mass energy; Plasma; Particle accelerator.

\bibliography{refs.bib}
\bibliographystyle{apsrev4-2}

\end{document}